\documentclass[11pt]{article}

\usepackage{amsmath,amsfonts,amssymb,amsthm,epsfig,epstopdf,titling,url,array}
\usepackage{graphicx, psfrag}
\usepackage{dynkin-diagrams}
\usepackage{appendix}
\usepackage{cite}
\usepackage{hyperref}
\usepackage[utf8]{inputenc}
\usepackage[english]{babel}
\usepackage{color}
\usepackage{booktabs,caption}
\usepackage[flushleft]{threeparttable}
\usepackage{lscape}
\definecolor{Mygrey}{gray}{0.8}
\definecolor{Mywhite}{gray}{1.0}

\newcommand{\be}{\begin{equation}}
\newcommand{\ee}{\end{equation}}
\newcommand{\bea}{\begin{eqnarray}}
\newcommand{\eea}{\end{eqnarray}}
\begin{document}

\begin{center}
\textbf{\Large \bf Wronskian Indices and Rational Conformal Field Theories}
\end{center}

\vskip .6cm
\medskip

\vspace*{4.0ex}

\baselineskip=18pt

\centerline{\large \rm   Arpit Das$^{1a}$, Chethan N. Gowdigere$^{2bc}$, Jagannath Santara$^{3bc}$}

\vspace*{4.0ex}

\centerline{\large \it $^a$ Centre for Particle Theory, Department of Mathematical Sciences,}

\centerline{\large \it  Durham University, South Road, Durham DH1 3LE, UK}

\vspace*{1.0ex}

\centerline{\large \it $^b$ National Institute of Science Education and Research Bhubaneshwar,}

\centerline{\large \it  P.O. Jatni, Khurdha, 752050, Odisha, INDIA}

\vspace*{1.0ex}

\centerline{\large \it $^c$ Homi Bhabha National Institute, Training School Complex, }

\centerline{\large \it  Anushakti Nagar, Mumbai 400094, INDIA}

\vspace*{4.0ex}
\centerline{E-mail: $^1$arpit.das@durham.ac.uk, $^2$chethan.gowdigere@niser.ac.in, $^3$jagannath.santra@niser.ac.in}

\vspace*{5.0ex}
\centerline{\bf Abstract} \bigskip
The classification scheme for rational conformal field theories, given by the Mathur-Mukhi-Sen (MMS) program, identifies a rational conformal field theory by two numbers: $(n, l)$. $n$ is the number of characters of the rational conformal field theory. The characters form linearly independent solutions to a modular linear differential equation (which is also labelled by $(n, l)$); the Wronskian index $l$ is a non-negative integer associated to the structure of zeroes of the Wronskian. 

In this paper, we compute the $(n, l)$ values for three classes of well-known CFTs viz. the WZW CFTs, the Virasoro minimal models and the $\mathcal{N} = 1$ super-Virasoro minimal models. For the latter two, we obtain exact formulae for the Wronskian indices. For WZW CFTs, we get exact formulae for small ranks (upto 2) and all levels and for all ranks and small levels (upto 2) and for the rest we compute using a computer program. We find that any WZW CFT at level 1 has a vanishing Wronskian index as does the $\mathbf{\hat{A}_1}$ CFT at all levels. We find intriguing coincidences such as: (i) for the same level CFTs with $\mathbf{\hat{A}_2}$ and $\mathbf{\hat{G}_2}$ have the same $(n,l)$ values, (ii) for the same level CFTs with $\mathbf{\hat{B}_r}$ and $\mathbf{\hat{D}_r}$ have the same $(n,l)$ values for all $r \geq 5$. 

Classifying all rational conformal field theories for a given $(n, l)$ is one of the aims of the MMS program. We can use our computations to provide partial classifications. For the famous $(2, 0)$ case, our partial classification turns out to be the full classification (achieved by MMS three decades ago). For the $(3, 0)$ case, our partial classification includes two infinite series of CFTs as well as fifteen ``discrete'' CFTs; except  three all others have Kac-Moody symmetry.

\vfill \eject

\baselineskip=18pt

\tableofcontents

\section{Introduction \label{1s}}

Two dimensional conformal field theory (CFT) is a very important subject since it bears relevance to many areas in physics and mathematics; \cite{DiFrancesco:1997nk, Moore:1989vd, Fuchs:2009iz, Gaberdiel:1999mc} are a partial list of references for CFTs. In physics it is important for string theory where it was intially relevant for the world-sheet description and for string perturbation theory, and later for various aspects in non-perturbative string theory also. Two dimensional CFTs are relevant for studying two dimensional critical systems at the fixed point of the renormalisation group. They are relevant for the Kondo problem, describe universality classes of quantum Hall fluids and find applications in entanglement entropy and quantum computing. Two dimensional CFTs have contributed to many areas of mathematics including representation theory, infinite dimensional algebras, theory of modular forms, etc.

An important class of two dimensional conformal field theories is that of  rational conformal field theories (RCFTs). In these theories the central charge $c$ and the conformal dimensions of the primary fields, the  $h$s, are all rational numbers \cite{Anderson:1987ge}. An infinite number of RCFTs are known to exist. One example is the infinite series of Virasoro minimal models \cite{Belavin:1984vu} (see also \cite{DiFrancesco:1997nk}), which are labelled by a pair of co-prime integers $(p, p')$. A second example is the infinite set of Wess-Zumino-Witten conformal field theories (WZW CFTs) \cite{Witten:1983ar}, whose spectrum-generating-algebra is an affine Lie algebra which have two pieces of data viz. a simple Lie algebra $\mathbf{g}$ and a positive integer $k$ known as the level. We will be studying these two classes of RCFTs in this paper. Even though an infinite number of RCFTs are known, there is no known complete classification of them. 

One classification scheme for RCFTs is based on the structure of their torus partition functions. For a RCFT, the torus partition function can be expressed as a sum of holomorphically factorised terms. The holomorphic factors are the characters of the RCFT and the number of characters, denoted by $n$, is an important detail of a RCFT for this classification scheme. Another important detail of a RCFT relevant for this classification scheme arises as follows. The $n$ characters are the linearly independent solutions of a single $n$-th order ordinary differential equation on the moduli space of a torus, a modular linear differential equation (MLDE) \cite{Eguchi:1987qd, Mathur:1988gt, Anderson:1987ge}. One then considers the Wronskian of these solutions which is a modular function of weight $n (n-1)$. The Wronskian index of the RCFT, denoted by $l$, is a number that takes non-negative integer values and is associated with the structure of zeroes of the Wronskian and can be expressed purely in terms of the RCFT data (the central charge, the number of characters and the conformal dimensions of the primary fields). In this classification scheme, a RCFT is identified in terms of these two numbers $(n, l)$, the number of characters and the Wronskian index. 

The coefficient functions of the MLDE are fixed by the pair $(n,l)$, apart from some undetermined parameters. Thus two RCFTs with the same $(n, l)$ values are both solutions to the same MLDE but for a different set of parameters in the MLDE.  One thus sets up a MLDE for a given number of characters and a given Wronskian index and studies all character-like solutions (that is solutions with non-negative integer coefficients in their $q$-expansion) and obtains all RCFTs with that number of characters and that Wronskian index and thus classifies them. This is the MLDE approach to the classification of RCFTs, first carried out in \cite{Mathur:1988na}, where a two-character vanishing-Wronskian-index i.e $(n = 2, l = 0)$ MLDE was studied and all two-character vanishing-Wronskian-index RCFTs were classified. This MLDE approach to RCFT is referred to as the MMS (Mathur-Mukhi-Sen) program. The program has been implemented for two-character RCFTs with Wronskian-index $2$, first in \cite{Naculich:1988xv} and then in \cite{Hampapura:2015cea, Gaberdiel:2016zke, Hampapura:2016mmz}; then for three-character vanishing-Wronskian-index RCFTs in \cite{Mathur:1988gt, Mukhi:2020gnj, Tener:2016lcn, kaneko4}. The MLDE approach to RCFTs has been covered in both the physics \cite{Hampapura:2015cea, Gaberdiel:2016zke, Mukhi:2020sxt, Bantay:2010uy, Tener:2016lcn, Harvey:2018rdc, Harvey:2019qzs, Bae:2017kcl, Bae:2020xzl}  and the maths literature \cite{kaneko5, kaneko1, kaneko2, kaneko3, Gannon:2013jua, arike1, franc1, kaneko4, mason1}. A status report of the program can be found in \cite{Mukhi:2019xjy}.

The MLDE approach to RCFTs, the MMS program, is also expected to discover new RCFTs, given that it is a systematic classification scheme. While this has happened in the literature \cite{Naculich:1988xv, Hampapura:2015cea, Gaberdiel:2016zke, Hampapura:2016mmz}. it also happens that many known RCFTs are rediscovered. For example, in \cite{Mathur:1988na}, nine solutions to the two-character vanishing-Wronskian-index MLDE were discovered. One of them turns out not to be associated with a CFT.  Each of the other eight solutions are associated to known RCFTs; one of them is the only two-character Virasoro minimal model with $(p = 5, p' = 2)$ and other seven are WZW CFTs with $\mathbf{g} = \mathbf{A_1}, \mathbf{A_2}, \mathbf{G_2}, \mathbf{D_4}, \mathbf{F_4}, \mathbf{E_6}, \mathbf{E_7}$, all at level $k = 1$.  One could perhaps ask the question: are there any new RCFTs with a given $(n, l)$ besides the known RCFTs such as the Virasoro minimal models, the WZW CFTs and others, and various tensor products of these theories? To ask this question, we would need to know the $(n, l)$ values of known RCFTs and this is one of the things we study in this paper: we compute the number of characters and the Wronskian-indices for a large class of RCFTs. We study the WZW CFTs for all Lie algebras and all levels and obtain exact formulae for many cases and for others via computer programs. We also study the Virasoro minimal models and obtain the Wronskian indices exactly and we do the same for the $\mathcal{N} = 1$ super-Virasoro minimal models. Our computations allow us to anticipate solutions for various MLDEs, and hence hopefully make it somewhat easier to solving these MLDEs.

This paper is organized as follows. In section \ref{2s}, we give a more detailed introduction to the MLDE approach to RCFTs, expanding on the brief version given above. In section \ref{3s}, we take up the study of WZW CFTs. We begin in \ref{31s} collecting details about WZW CFTs and outlining a procedure to compute the $n$ and $l$ value for any WZW CFT. In \ref{32s} to \ref{35s}, we successively study the WZW CFTs based on the Lie algebras $\mathbf{A}$ , $\mathbf{B}$, $\mathbf{C}$, $\mathbf{D}$ and the exceptional series.  Where we can, we obtain exact formulae for an arbitrary level;  we can do this is for all rank two Lie algebras $\mathbf{A_2}$, $\mathbf{C_2}$ and $\mathbf{G_2}$. For other ranks, we use a computer program and tabulate the results (see tables in appendix \ref{app1}). From the tables we notice that all level one WZW CFTs seem to have a vanishing Wronskian index which we are able to prove by an exact computation in \ref{37s}. We also note some patterns about level two WZW CFTs  which we are able to prove by exact computations in \ref{38s}. In  \ref{39s}, we summarise the results and provide various conjectures about WZW CFTs. In section \ref{4s}, we first study Virasoro minimal models in \ref{41s} and are able to perform an exact computation for the infinite series of RCFTs and remarkably, they all have a vanishing Wronskian index.  Later in \ref{42s} we study the $\mathcal{N} = 1$ super-Virasoro minimal models and we obtain an exact expression for this infinite series of RCFTs as well. Finally, in section \ref{5s} we conclude by discussing the main results of the paper and directions for future work.

\section{MLDE Approach to RCFTs \label{2s}}
The characters of a RCFT are a set of $n$ functions, $\chi_i(\tau)$, on the moduli space of the torus such that the torus partition function of the RCFT can be written as: 
\be \label{1n}
Z(\tau,\bar{\tau})=\sum_{i,j=1}^{n} M_{ij}\chi_i(\tau)\chi_j(\bar{\tau})
\ee
The characters have a $q$-expansion:
\be \label{2n}
\chi_i(\tau)=\sum_{n=0}^\infty a_n^{(i)} q^{\alpha_i+n},\qquad i = 1,\cdots,n-1
\ee
where $q=e^{2\pi i\tau}$. Here $\alpha_i$ are the exponents, corresponding to $h_i-\frac{c}{24}$ where $h_i$ are the conformal dimensions of the primaries and $c$ is the central charge. Note that, a single character can correspond to multiple primaries. The torus partition function is required to be modular invariant:
\be \label{3n}
Z(\gamma \tau,\gamma\bar{\tau}) = Z(\tau,\bar{\tau}),\quad \gamma=
\begin{pmatrix}
a~ & ~b~\\ c~ & ~d~
\end{pmatrix}
\in SL(2,\mathbf{Z})
\ee
This is ensured if the characters transform as vector-valued modular forms:
\be \label{4n}
\chi_i(\gamma\tau)= \sum_kV_{ik}(\gamma)\chi_k(\tau)
\ee
where $V_{ik}$ are unitary matrices in the $n$-dimensional representation of $SL(2, \mathbf{Z})$.

\subsection{Modular Linear Differential Equations \label{21s}}
\subsubsection{The Serre-Ramanujan derivative operator \label{211}}
We denote the upper half plane $\mathbb{H}$ as,
\begin{align} \label{5n}
\mathbb{H} = \{\tau\in\mathbb{C} \ | \ \text{Im}(\tau)>0\}. 
\end{align}
The {\it Serre-Ramanujan} derivative operator (see, for example, section 2.8 of \cite{kilford} and exercise 5.1.8 of \cite{rammurthy1}) is defined as follows:
\begin{align} \label{6n}
\mathcal{D} := \partial_\tau - \frac{\iota\pi k}{6}E_2(\tau), 
\end{align}
where $E_2(\tau)$ is the weight $2$ Eisenstein series. The operator $\mathcal{D}$  is a linear operator which maps weight $m$ modular objects to weight $m+2$ modular objects and satisfies the Leibniz rule. This is the differential operator with which one sets up the modular linear differential equations and defines the associated Wronskians.

\subsubsection{$n^{th}$ Order MLDE \label{212}}
With this we can write the most general $n^{th}$ order MLDE \cite{Mathur:1988na}:
\begin{align}
\mathcal{D}^n \chi_i + \sum_{r=0}^{n-1} \phi_r(\tau)\mathcal{D}^r \chi_i = 0, \label{7n}    
\end{align}
where the solutions $\chi_i$ denote the characters of a $n$-character RCFT. Note that $n$ denotes the number of linearly independent characters in a RCFT. We seek for Frobenius kind of solutions to the above MLDE,
\begin{align}
\chi_i = q^{\alpha_i}\sum_{n=0}^\infty \chi_{i,n} q^n, \label{8n}    
\end{align}
with the weight $k$ Eisenstein series $q$-expansion given by,
\begin{align}
E_k(q) = \sum_{n=0}^\infty E_{k,n} q^n. \label{9n}   
\end{align}
The Eisenstein series is normalised such that $E_{k,0}=1$. The exponents $\alpha_i$ are related to the central charge $c$ and conformal dimensions $h_i$ of the theory, as follows,
\begin{align}
\alpha_i = -\frac{c}{24} + h_i, \label{10n}
\end{align}
Furthermore, since $\chi_i$ denote characters of a RCFT, their coefficients, that is, $\chi_{i,n}$ have to be non-negative integers.

\subsubsection{The Wronskian $W$}
For the differential equation \eqref{7n}, define,
\begin{align}
W_r = \left(
\begin{array}{cccc}
    \chi_1 & \cdots & \cdots & \chi_{n} \\
    \chi_1^{(1)} & \cdots & \cdots & \chi_{n}^{(1)} \\
    \vdots & \ddots & \ddots & \vdots \\
    \chi_1^{(r-1)} & \cdots & \cdots & \chi_{n}^{(r-1)} \\
    \chi_1^{(r+1)} & \cdots & \cdots & \chi_{n}^{(r+1)} \\
    \vdots & \ddots & \ddots & \vdots \\
    \chi_1^{(n)} & \cdots & \cdots & \chi_{n}^{(n)}
\end{array}
\right), \label{11n}
\end{align}
where $\chi_i^{(m)} \equiv \mathcal{D}^m \chi_i$. The Wronskian, $W$, of \eqref{11n} is given by $W\equiv W_n$. After some computations one can get the following relation,
\begin{align}
\phi_r = (-1)^{n-r}\frac{W_r}{W}. \label{phir}    
\end{align}
Now let us analyse the weights of various objects in \eqref{7n}. $\chi_i$s being characters of a RCFT are weight $0$ modular functions. The first term, $\mathcal{D}^n \chi_i$, is of weight $2n$ as each operation of $\mathcal{D}$ increases the weight by $2$. Similarly  $\mathcal{D}^r \chi_i$ is of weight $2r$ and hence $\phi_r(\tau)$ are modular functions of weight $2(n-r)$. Note that since $W$ can have zeros in $\mathbb{H}$, so $\phi_r$s are not modular forms but rather are modular functions\footnote{Modular forms are holomorphic in $\mathbb{H}$ but modular functions, in general, are meromorphic in $\mathbb{H}$.}. Also, note that $W$ is a modular function of weight $n(n-1)$.

\subsubsection{The Valence Formula and the Wronskian Index}
For a modular function $f$ of weight $k$, $\nu_{\tau_0}(f)$ denotes the order of pole/zero of $f$ at $\tau=\tau_0$. Let us now look at the valence formula (see, for example, section 3.1 of \cite{kilford}) which states that,
\begin{align}
\nu_{\iota\infty}(f) + \frac{1}{2}\nu_\iota(f) + \frac{1}{3}\nu_\omega(f) + \sum^{'}_{\tau\neq \iota,\omega; \ \tau\in\mathcal{F}} \nu_\tau(f) = \frac{k}{12}, \label{VF}   
\end{align}
where $\omega$ is one of the cube root of unity, $\mathcal{F}$ is the Fundamental Domain and the $'$ above the summation means that the sum excludes $\tau\in\mathbb{H}$ which have $\text{Re}(\tau)=\frac{1}{2}$ or, which have both $|\tau|=1$ and $\text{Re}(\tau)>0$.\\
\\
Now, at $\tau\rightarrow\iota\infty$, $\chi_i\sim q^{\alpha_i}$ and hence, $W\sim q^{\sum_{i=0}^{n-1}\alpha_i}$. Hence, $W$ has a $-\sum_{i=0}^{n-1}\alpha_i$ order pole at $\iota\infty$. The last three terms on the left hand side of \eqref{VF} together can be written as  $\frac{l}{6}$ where $l$ is a non-negative integer. Hence, the valence formula now reads
\begin{align}
\sum_{i=1}^{n}\alpha_i + \frac{l}{6} = \frac{n(n-1)}{12}  . \label{RR}    
\end{align}
Now using \eqref{10n}, we have 
\be \label{windex}
\frac{l}{6} = \frac{n(n - 1)}{12} + \frac{n\,c}{24} - \sum_i h_i.
\ee
$l$, which is some information about the zeroes of the Wronskian (at the cusp points and the interior of the fundamental domain), is the Wronskian index of the RCFT. This Wronskian index $l$, which is expressed purely in terms of the RCFT data (as seen in \eqref{windex}), proves to be important for the classification of RCFTs.

\section{Wess-Zumino-Witten CFTs \label{3s}}

\subsection{Generalities \label{31s}}
To every Wess-Zumino-Witten CFT, there is an affine Lie algebra at a certain level viz $\hat{\mathbf{g}}_k$, which is it's spectrum generating algebra. The ingredients of the affine Lie algebra are a finite simple Lie algebra $\mathbf{g}$ and a positive integer $k$, called the level.  If the finite simple Lie algebra has rank $r$, it has $r$ dimensional Cartan subalgebra and $r$ dimensional dual of the Cartan subalgebra which is where the root lattice and the weight lattices live. There are the usual preferred basis of this dual of the Cartan subalgebra viz. the simple root basis, the co-root basis and the fundamental weight basis. There is a Dynkin diagram with $r$ nodes associated with the simple Lie algebra. 
The affine Lie algebra based on a finite Lie algebra of rank $r$ can be thought of in terms of a Dynkin diagram with $r+1$ nodes. There is one node (the zeroth) added to the $r$ nodes of the Dynkin diagram of the finite algebra $\mathbf{g}$. All the information for simple Lie algebras that we will need are collected in an appendix \ref{app2}.

The highest weight representations of an affine lie algebra  $\hat{\mathbf{g}}_k$ are defined by their highest weights which are $[\lambda_0, \lambda_1, \ldots, \lambda_r]$, with  $r+1$ Dynkin labels, all of which are non-negative integers.  The last $r$ labels define a highest weight representation of the finite lie algebra $\mathbf{g}$. 

An important class of representations relevant for two dimensional CFT are the integrable highest weight representations. These correspond to WZW primary fields. Integrable highest weight representations are those whose highest weights have Dynkin labels restricted (by the level) to satisfy the equation: 
\be \label{level-label}
k = \sum_{i = 0}^{r} \check{a}_i\,\lambda_i = \lambda_0 + \sum_{i = 1}^{r} \check{a}_i\,\lambda_i
\ee
where $\check{a}_i$ are the co-marks of the Lie algebra. The central charge of a Wess-Zumino-Witten CFT for $\hat{\mathbf{g}}_k$ is given by:
\be \label{centralcharge}
c = \frac{k ~\text{dim}\, \mathbf{g}}{k + \mathit{g}}
\ee
where $\mathit{g}$ is the dual Coxeter number of the finite algebra $\mathbf{g}$. The primary field associated with the integral highest weight representation with highest weight $[\lambda_0, \lambda_1, \ldots, \lambda_r]$, restricted by \eqref{level-label}, has a conformal dimension given by
\be \label{h}
h_{\lambda} = \frac{( \lambda, \lambda + 2 \rho)}{2 (k + \mathit{g})}
\ee
where $\lambda$ is $[\lambda_1, \ldots, \lambda_r]$, the highest weight representation of $\mathbf{g}$. Here $\rho$ is the Weyl vector, which is a weight of the finite algebra $\mathbf{g}$, with all it's Dynkin labels equal to $1$. The inner-product in the numerator of \eqref{h} is given by 
\be \label{5} 
(\lambda, \mu) = \sum_{i,j=1}^{r}\lambda_i \,\mu_j\, F_{ij}
\ee
where $F_{ij}$ is the quadratic form matrix of the finite algebra $\mathbf{g}$.

To compute the Wronskian index of a WZW CFT, we first need to compute the number of characters. This is (roughly) the number of WZW primary fields, that is, the number of integrable highest weight representations of the affine Lie algebra. We will mostly refer to the WZW primary field by the $\mathbf{g}$ representation $[\lambda_1, \cdots \lambda_r ] $ associated with it. The understanding is that the Virasoro algebra belongs to the universal enveloping algebra of the affine algebra and hence it suffices to classify the Virasoro primary fields in terms of the irreducible representations of the affine Lie algebra.  The number of Virasoro primary fields in WZW CFTs may be infinite but the number of WZW primary fields (w.r.t. the affine Lie algebra) and hence the number of associated characters is finite (see Section 15.3.5 of \cite{DiFrancesco:1997nk}). Hence to get the number of characters of a WZW CFT, we find all non-negative solutions to \eqref{level-label}, modulo any extra symmetries of the Dynkin diagram. Then we compute the associated conformal dimensions \eqref{h} and together with the central charge of the theory \eqref{centralcharge}, we obtain the Wronskian index of the WZW CFT \eqref{windex}. The procedure that we have just outlined reproduces, among others,  the computations of \cite{Mathur:1988na} etc., and we follow this in the rest of this paper. From here on, in the rest of this section \ref{3s}, we shall refer to WZW primary fields as just primary fields.
 
 \subsection{$\mathbf{A}$ series \label{32s}}
\subsubsection{$(\mathbf{\hat{A}_1})_k$ \label{321}}
For the $(\mathbf{\hat{A}_1})_k$ CFT, the affine algebra has rank two and the finite algebra has rank one;  the dimension of the algebra is $3$ and the dual Coxeter number is $2$ (equation \eqref{2aDG} of appendix \ref{app2}), which when used in \eqref{centralcharge} gives the central charge:
\be \label{6}
c \,[\mathbf{\hat{A}_1}; k] = \frac{3\,k}{k +2}.
\ee
From the Dynkin diagram of $(\mathbf{\hat{A}_r})_k$ (see appendix \ref{app2}), we note that all the co-marks are 1 and hence the integrable representations,  $[\lambda_0, \lambda_1]$, are restricted by \eqref{level-label}:
\be \label{7}
k = \lambda_0 + \lambda_1,
\ee
which further means that the CFT has $k+1$ primary fields in the $[0], [1], \ldots [k]$ representations of $A_1$. 
\be \label{8}
n \,[\mathbf{\hat{A}_1}; k] = k + 1
\ee
To  compute the conformal dimensions of these primary fields, first we note the quadratic form matrix, which is a $1 \times 1$ matrix, is the number $\frac12$ so that  $([\lambda], [\mu]) = \frac12 \,\lambda\, \mu $
and we obtain from \eqref{h}
\be \label{9}
h_{[\mu]} = \frac{\mu (\mu + 2)}{4 (k + 2)}.
\ee
We need the sum of the scaling dimensions of the primary fields, which can be computed using high school formulae, $\sum\limits_{\mu = 0}^{k} \mu = \frac{k (k+1)}{2}, ~ \sum\limits_{\mu = 0}^{k} \mu^2 = \frac{k (k+1) (2 k + 1)}{6}$:
\be \label{10}
\sum_i h_i = \sum_{\mu = 0}^{k} \frac{\mu (\mu + 2)}{4 (k + 2)} = \frac{k (k+1) (2 k + 7)}{24 (k+2)}.
\ee
We can then assemble the  equations \eqref{6}, \eqref{8}, \eqref{10} to compute the Wronskian index \eqref{windex} of the  $(\mathbf{\hat{A}_1})_k$ CFT:
\be \label{11}
l\,[\mathbf{\hat{A}_1}; k] = 0.
\ee
We have thus obtained the result that for all levels, the $(\mathbf{\hat{A}_1})_k$ CFTs have a vanishing Wronskian index. We will see that this vanishing of the Wronskian index is more of an exception than a rule. We note (from \eqref{8}) that the $(\mathbf{\hat{A}_1})_k$ is a two-character vanishing-Wronskian-index CFT; this appears as one of the solutions in \cite{mathur1} where a study of two-character vanishing-Wronskian-index MLDE was first made. 
\subsubsection{$(\mathbf{\hat{A}_2})_k$ \label{322}}

For the $(\mathbf{\hat{A}_2})_k$ CFT, the affine algebra has rank three and the finite algebra has rank two; the dimension of the algebra is $8$ and the dual Coxeter number is $3$ (equation \eqref{2aDG} of appendix \ref{app2}), which when used in \eqref{centralcharge} gives the central charge:
\be \label{12}
c \,[\hat{A}_2; k] = \frac{8\,k}{k +3}.
\ee
From the Dynkin diagram of $(\mathbf{\hat{A}_r})_k$ (see appendix \ref{app2}), we note that all the co-marks are 1 and hence the integrable representations,  $[\lambda_0, \lambda_1, \lambda_2]$, are restricted by \eqref{level-label}:
\be \label{13}
k = \lambda_0 + \lambda_1 + \lambda_2.
\ee
For every non-negative integer solution to \eqref{13}, there is a primary field in the CFT in the $[\lambda_1, \lambda_2]$ representation whose scaling dimension, we obtain from \eqref{h}, is given by  
 \be \label{14}
h_{[\lambda_1, \lambda_2]} = \frac{2 \lambda_1^2 + 2 \lambda_1\lambda_2 + 2 \lambda_2^2 + 6 \lambda_1 + 6 \lambda_2}{6 (k + 3)}.
\ee 
Since the number $n$ that enters the formula for the Wronskian index \eqref{windex} is the number of primary fields with linearly independent characters, we need to identify representations whose characters are not linearly independent. This would happen when there is a symmetry in the Dynkin diagram as is the case in the Dynkin diagram of $\mathbf{A_2}$ where there is the following symmetry:
\be \label{15}
\lambda_1 \leftrightarrow \lambda_2,
\ee
which takes a $\mathbf{A_2}$ representation $[\lambda_1, \lambda_2]$ to it's complex conjugate representation  $[\lambda_2, \lambda_1]$ and one expects their characters to be not linearly independent\footnote{Note that this in particular means, using \eqref{14}, that $h_{[\lambda_1, \lambda_2]} = h_{[\lambda_2, \lambda_1]}.$  However, the converse is not true, that is, two representations with the same scaling dimensions may be linearly independent. For example, we find for levels $k \geq 9$, the representations $[5,4]$ and $[8,0]$ have the same scaling dimensions but linearly independent characters. Similar pairs of representations are  $[8,3]$ and $[10,0]$ for levels $k \geq 11$ and $[10,1]$ and $[6,6]$ for levels $k \geq 12$.}. Hence we need to count the number of solutions to \eqref{13} modulo the symmetry \eqref{15}.

At level $k$, there is one primary field for every partition of every non-negative integer less than or equal to $k$, into atmost two summands. For example, at level $1$, the partitions are of $0$ and $1$ into atmost two summands which are $0+0, 1+0$ respectively, the primary fields are in representations $[0,0], [1,0]$, that is, $n \,[\mathbf{\hat{A}_2}; 1] = 2$. Similarly, at level $2$, the partitions are of $0, 1$ and $2$ into atmost two summands which are $0+0, 1+0, 1+1, 2+0$ respectively, the primary fields are in representations $[0,0], [1,0], [1,1], [2,0]$, that is, $n \,[\mathbf{\hat{A}_2}; 2] = 4$. It is possible to count the number of partitions into atmost two summands and obtain the following formula for the number of primary fields:
\bea \label{16}
n \,[\mathbf{\hat{A}_2}; k] = & \frac{(k+2)^2}{4}, \qquad \text{for even}~ k \nonumber \\
& \frac{(k+1)(k+3)}{4} \qquad \text{for odd}~ k.
\eea
If we define
\be \label{17}
D_{[\lambda_1, \lambda_2]} \equiv 2 \lambda_1^2 + 2 \lambda_1\lambda_2 + 2 \lambda_2^2 + 6 \lambda_1 + 6 \lambda_2, \qquad 
D(m) \equiv \sum_{\lambda_1 + \lambda_2 = m} D_{[\lambda_1, \lambda_2]},
\ee
where the summation is over the partitions of $m$ into atmost two summands, then the summation over conformal dimensions of the primary fields can be written:
\be \label{18}
\sum_i h_i \equiv \frac{H(k)}{6 (k + 3)}, \qquad H(k) =  \sum_{m=0}^k D(m).
\ee
Now a formula for $D(m)$ can be obtained as
\bea 
D(m) &=& \sum_{n = 0}^{\frac{m}{2}} D_{[n, m-n]} = \sum_{n = 0}^{\frac{m}{2}} [2 m^2 + 6 m - 2 m n + 2 n^2] = \frac{m\, (m + 2) (10 m + 37)}{12}~~\qquad \text{for even}~ m \nonumber \\
&=& \sum_{n = 0}^{\frac{m-1}{2}} D_{[n, m-n]} = \sum_{n = 0}^{\frac{m-1}{2}} [2 m^2 + 6 m - 2 m n + 2 n^2] = \frac{m\, (m + 1) (10 m + 38)}{12}~~\qquad \text{for odd}~ m, \nonumber
\eea
from which one obtains the formula for $H(k)$:
\bea \label{19}
H(k) = & \frac{k (k + 2) (5 k^2 + 35 k + 48)}{24}, \qquad \text{for even}~ k \nonumber \\
& \frac{(k + 1) (k + 3) (5 k^2 + 25 k - 6)}{24} \qquad \text{for odd}~ k.
\eea
We can now assemble the equations \eqref{12}, \eqref{16}, \eqref{18}, \eqref{19} to compute the Wronskian index \eqref{windex} of the  $(\mathbf{\hat{A}_2})_k$ CFT:
\bea \label{20}
l \,[\mathbf{\hat{A}_2}; k] = & \frac{k (k^2 - 4) (3 k + 4)}{96}, \qquad \text{for even}~ k \nonumber \\
& \frac{ (k^2 - 1) (k + 3) (3 k - 5)}{96} \qquad \text{for odd}~ k.
\eea
We can see that the only theories with a vanishing Wronskian index are for $k = 1$ and $k=2$, both of which are two-character theories; and hence should (and does) appear in a study of two-character vanishing-Wronskian-index MLDEs \cite{Mathur:1988na}.  For all higher levels $k \geq 3$, the Wronskian index is non-zero. 

\subsubsection{$(\mathbf{\hat{A}_r})_k$ \label{323}}
For the $(\mathbf{\mathbf{\hat{A}_r}})_k$ CFT, the affine algebra has rank $r+1$ and the finite algebra has rank $r$; the dimension of the algebra is $r^2 + 2r$ and the dual Coxeter number is $r+1$ (equation \eqref{2aDG} of appendix \ref{app2}), which when used in \eqref{centralcharge} gives the central charge:
\be 
\label{21}
c \,[\mathbf{\hat{A}_r}; k] = \frac{k\,r\,(r+2)}{k + r + 1}.
\ee
From the Dynkin diagram of $(\mathbf{\hat{A}_r})_k$ (see appendix \ref{app2}), we note that all the co-marks are 1 and hence the integrable representations,  $[\lambda_0, \lambda_1, \ldots \lambda_r]$, are restricted by \eqref{level-label}:
\be \label{22}
k = \lambda_0 + \lambda_1 + \ldots + \lambda_r.
\ee
There is a symmetry in the Dynkin diagram of $A_r$:
\be \label{23}
\lambda_i \leftrightarrow \lambda_{r - i +1}
\ee
and two representations related by this symmetry are expected to have characters not linearly independent. Hence to obtain $n \,[\mathbf{\hat{A}_r}; k]$ we need to count the number of non-negative solutions to \eqref{22} modulo the symmetry \eqref{23}. We do this and the further computation of the scaling dimensions and eventually the Wronskian index via a computer program and the results are tabulated in table 1, for levels upto $12$, in three columns: one for the number of characters, one for the central charge and the last for the Wronskian index. The table does not include rank one for which the entire Wronskian index column would have been zero. We observe that for each rank, at level one, the Wronskian index vanishes. This motivates the question if this is true for all ranks and we will give an exact computation in subsequent sections for the vanishing of the Wronskian index for all $(\mathbf{\hat{A}_r})_1$. We also note from table 1 that at level $2$, only $(\mathbf{\hat{A}_2})_2$ (and $(\mathbf{\hat{A}_1})_2$ which is not shown in the table) has a vanishing Wronskian index; in subsequent sections, we will derive an exact formula for the Wronskian index of all $(\mathbf{\hat{A}_r})_2$ CFTs which will confirm the results in the table (only the second row). We also note that there are no vanishing-Wronskian-index CFTs apart from the level one CFTs ($(\mathbf{\hat{A}_r})_1$), the rank one CFTs ($(\mathbf{\hat{A}_1})_k$) and $(\mathbf{\hat{A}_2})_2$; all other CFTs have a non-zero Wronskian index. We can also use the table to anticipate solutions of MLDEs; for example, we can predict two solutions (for two different sets of parameters) for the three-character vanishing-Wronskian-index MLDE viz. the $(\mathbf{\hat{A}_3})_1$ and $(\mathbf{\hat{A}_4})_1$ CFTs.

\subsection{$\mathbf{B}$ series \label{33s}}
\subsubsection{$(\mathbf{\hat{B}_r})_k$ \label{331}}

For the $(\mathbf{\hat{B}_r})_k$ CFT, the affine algebra has rank $r+1$ and the finite algebra has rank $r$; the dimension of the algebra is $2 r^2 + r$ and the dual Coxeter number is $2 r-1$ (equation \eqref{2bDG} of appendix \ref{app2}), which when used in \eqref{centralcharge} gives the central charge:
\be \label{24}
c \,[\mathbf{\hat{B}_r}; k] = \frac{k\,r\,(2r+1)}{k + 2r - 1}.
\ee
From the Dynkin diagram of $(\mathbf{\hat{B}_r})_k$ (see appendix \ref{app2}),  we note that all the co-marks of the finite algebra are $2$ except the first and the last which is $1$ and hence the integrable representations,  $[\lambda_0, \lambda_1, \ldots \lambda_r]$, are restricted by \eqref{level-label}:
\be \label{25}
k = \lambda_0 + \lambda_1 + 2 \lambda_2 + 2 \lambda_3+ \ldots \ldots + 2 \lambda_{r-1} +  \lambda_r.
\ee
We obtain the number of primary fields $n \,[\mathbf{\hat{B}_r}; k]$  by counting the number of non-negative solutions to \eqref{25} via a computer program which also computes the scaling dimensions and  eventually the Wronskian index and the results are tabulated in table 2, for levels upto $12$.  We again observe that for each rank, at level one, the Wronskian index vanishes. This again motivates the question if this is true for all ranks and we will give an exact computation in subsequent sections for the vanishing of the Wronskian index for all $(\mathbf{\hat{B}_r})_1$. At level two, we note that the three quantities $n \,[\mathbf{\hat{B}_r}; 2]$, $c \,[\mathbf{\hat{B}_r}; 2]$ and $l \,[\mathbf{\hat{B}_r}; 2]$ seem to form three arithmetic progressions (constant difference between consecutive elements). We will be able to derive an exact formula for the Wronskian index of all $(\mathbf{\hat{B}_r})_2$ CFTs also which will confirm this observation.  We also note that there are no vanishing-Wronskian-index CFTs apart from the level one CFTs, $(\mathbf{\hat{B}_r})_1$. We can again use the table to anticipate solutions of MLDEs; for example, we can predict four solutions (for four different sets of parameters) for the three-character vanishing-Wronskian-index MLDE viz. the $(\mathbf{\hat{B}_3})_1, (\mathbf{\hat{B}_4})_1, (\mathbf{\hat{B}_5})_1$ and $(\mathbf{\hat{B}_6})_1$ CFTs.

\subsection{$\mathbf{C}$ series \label{34s}}
\subsubsection{$(\mathbf{\hat{C}_2})_k$ \label{341}}

For the $(\mathbf{\hat{C}_2})_k$ CFT, the affine algebra has rank two and the finite algebra has rank one;  the dimension of the algebra is $10$ and the dual Coxeter number is $2$ (equation \eqref{2cDG} of appendix \ref{app2}), which when used in \eqref{centralcharge} gives the central charge:
\be \label{28}
c \,[\mathbf{\hat{C}_2}; k] = \frac{10\, k}{k + 3}.
\ee
From the Dynkin diagram of $(\mathbf{\hat{C}_r})_k$ (see appendix \ref{app2}), we note that all the co-marks are 1 and hence the integrable representations,  $[\lambda_0, \lambda_1, \lambda_2]$, are restricted by \eqref{level-label}:
\be \label{29}
k = \lambda_0 + \lambda_1 + \lambda_2.
\ee
For every non-negative integer solution to \eqref{29}, there is a primary field in the CFT in the $[\lambda_1, \lambda_2]$ representation. In the absence of any symmetry in the Dynkin diagram, we expect the number of solutions to also be the number $n$ that enters into the formula \eqref{windex}, which we can obtain explicitly:
\be \label{30}
n \,[\mathbf{\hat{C}_2}; k] =  \frac{(k+2)(k+1)}{2}.
\ee
The conformal dimension of the primary field in the $[\lambda_1, \lambda_2]$ representation, we obtain from \eqref{h}, is given by  
 \be \label{31}
h_{[\lambda_1, \lambda_2]} = \frac{ \lambda_1^2 + 2 \lambda_1\lambda_2 + 2 \lambda_2^2 + 4 \lambda_1 + 6 \lambda_2}{4 (k + 3)}.
\ee 
If we define
\be \label{32}
D_{[\lambda_1, \lambda_2]} \equiv \lambda_1^2 + 2 \lambda_1\lambda_2 + 2 \lambda_2^2 + 4 \lambda_1 + 6 \lambda_2, \qquad 
D(m) \equiv \sum_{\lambda_1 + \lambda_2 = m} D_{[\lambda_1, \lambda_2]},
\ee
then the summation over conformal dimensions of the primary fields can be written:
\be \label{33}
\sum_i h_i \equiv \frac{H(k)}{4 (k + 3)}, \qquad H(k) =  \sum_{m=0}^k D(m).
\ee
Now a formula for $D(m)$ can be obtained as
\be \label{34}
D(m) = \sum_{n = 0}^{m} D_{[n, m-n]} = \frac{m\, (m + 1) (8 m + 31)}{6} 
\ee
from which one obtains the formula for $H(k)$:
\be \label{35}
H(k) =  \frac{k (k + 1) (k + 2) ( 2 k + 11)}{6}.
\ee
We can now assemble the equations \eqref{28}, \eqref{30}, \eqref{33}, \eqref{35} to compute the Wronskian index \eqref{windex} of the  $(\mathbf{\hat{C}_2})_k$ CFT:
\be \label{36}
l \,[\mathbf{\hat{C}_2}; k] =  \frac{( k - 1) k  (k + 1) (k + 2)}{8},
\ee
which is guaranteed to be a whole number since the numerator being a product of four consecutive integers  always has $4!$ as a factor. 

We can see, from \eqref{36}, that the only theory with a vanishing Wronskian index is the $k = 1$ theory which is a  three-character theory; and hence should  appear in a study of three-character vanishing-Wronskian-index MLDEs.  For all higher levels $k \geq 2$, the Wronskian index is non-zero. 

\subsubsection{$(\mathbf{\hat{C}_r})_k$ \label{342}}

For the $(\mathbf{\hat{C}_r})_k$ CFT, the affine algebra has rank $r+1$ and the finite algebra has rank $r$; the dimension of the algebra is $2 r^2 + r$ and the dual Coxeter number is $r+1$ (equation \eqref{2cDG} of appendix \ref{app2}), which when used in \eqref{centralcharge} gives the central charge:
\be 
\label{37}
c \,[\hat{C}_r; k] = \frac{k\,r\,(2 r + 1)}{k + r + 1}.
\ee
From the Dynkin diagram of $(\mathbf{\hat{C}_r})_k$ (see appendix \ref{app2}), we note that all the co-marks are 1 and hence the integrable representations,  $[\lambda_0, \lambda_1, \ldots \lambda_r]$, are restricted by \eqref{level-label}:
\be \label{38}
k = \lambda_0 + \lambda_1 + \ldots + \lambda_r.
\ee
We can obtain a formula for the number of primary fields
\be
n \,[\mathbf{\hat{C}_r}; k] = \frac{(k + r )!}{k ! ~ r !},
\ee
which is a degree $r$ polynomial in $k$ and we can expect the Wronskian index to be a degree $2r$ polynomial in $k$. We will not pursue the exact computation here. We  compute the scaling dimensions and   the Wronskian index via a computer program and the results are tabulated in table 3, for levels upto $12$. We again observe that for each rank, at level one, the Wronskian index vanishes. This again motivates the question if this is true for all ranks and we will give an exact computation in subsequent sections for the vanishing of the Wronskian index for all $(\mathbf{\hat{C}_r})_1$. We will also able to derive an exact formula for the Wronskian index of all $(\mathbf{\hat{C}_r})_2$ CFTs. We also note that there are no vanishing-Wronskian-index CFTs apart from the level one CFTs, $(\mathbf{\hat{C}_r})_1$. We can again use the table to anticipate solutions of MLDEs; for example, we can predict a solution for the three-character vanishing-Wronskian-index MLDE viz. the $(\mathbf{\hat{C}_2})_1$ CFT.

\subsection{$\mathbf{D}$ series \label{35s}}

\subsubsection{$(\mathbf{\hat{D}_4})_k$ \label{351}}

For the $(\mathbf{\hat{D}_4})_k$ CFT, the affine algebra has rank $5$ and the finite algebra has rank $4$; the dimension of the algebra is $28$ and the dual Coxeter number is $6$ (equation \eqref{2dDG} of appendix \ref{app2}), which when used in \eqref{centralcharge} gives the central charge:
\be \label{38a}
c \,[\mathbf{\hat{D}_4}; k] = \frac{28\,k}{k + 6}.
\ee
From the Dynkin diagram of $(\mathbf{\hat{D}_r})_k$ (see appendix \ref{app2}), we note that all the co-marks of the finite algebra are $1$ except the second one which is  $2$ and hence the integrable representations,  $[\lambda_0, \lambda_1, \ldots \lambda_4]$, are restricted by \eqref{level-label}:
\be \label{39a}
k = \lambda_0 + \lambda_1 + 2 \lambda_2  +  \lambda_3 + \lambda_4.
\ee
There is a symmetry in the Dynkin diagram of $\mathbf{D_4}$, the triality symmetry generated by
\be \label{40a}
\lambda_1 \rightarrow \lambda_2, \quad \lambda_2 \rightarrow \lambda_3, \quad \lambda_3 \rightarrow \lambda_1,
\ee
We obtain the number of primary fields by counting the number of non-negative solutions to \eqref{39a} modulo the symmetry \eqref{40a}. We do this via a computer program also obtaining the scaling dimensions and the Wronskian index; the results can be found in table 4 upto level $12$. We note  that the $(\mathbf{\hat{D}_4})_1$ is a two-character vanishing-Wronskian-index CFT; this appears as one of the solutions in \cite{Mathur:1988na} where a study of two-character vanishing-Wronskian-index MLDE was first made. We also note that there is a vanishing-Wronskian index CFT at level two, which is a five character theory and at all higher levels, the CFTs have a non-vanishing Wronskian index. 

\subsubsection{$(\mathbf{\hat{D}_r})_k$ \label{352}}

For the $(\mathbf{\hat{D}_r})_k$ CFT, the affine algebra has rank $r+1$ and the finite algebra has rank $r$; the dimension of the algebra is $2 r^2 - r$ and dual Coxeter number is $2 r-2$ (equation \eqref{2dDG} of appendix \ref{app2}), which when used in \eqref{centralcharge} gives the central charge:
\be \label{42}
c \,[\mathbf{\hat{D}_r}; k] = \frac{k\,r\,(2r-1)}{k + 2r - 2}.
\ee
From the Dynkin diagram of $(\mathbf{\hat{D}_r})_k$ (see appendix \ref{app2}), we note that all the co-marks of the finite algebra are $2$ except the first and the last two which are all $1$ and hence the integrable representations,  $[\lambda_0, \lambda_1, \ldots \lambda_r]$, are restricted by \eqref{level-label}:
\be \label{43}
k = \lambda_0 + \lambda_1 + 2 \lambda_2 + 2 \lambda_3+ \ldots \ldots + 2 \lambda_{r-2} +   \lambda_{r-1} +  \lambda_r.
\ee
Here in this subsection, we study the cases $r \geq 5$ for which there is no triality symmetry in the Dynkin diagram, but there is a symmetry:
\be \label{43a}
\lambda_{r-1} \leftrightarrow \lambda_r.
\ee 
We obtain the number of primary fields by counting the number of non-negative solutions to \eqref{43} modulo the symmetry \eqref{43a} via the computer program which also computes the scaling dimensions and  the Wronskian index;  the results are tabulated in table 4, for levels upto $12$.  We again observe that for each rank, at level one, the Wronskian index vanishes. We will give an exact computation in subsequent sections for the vanishing of the Wronskian index for all $(\mathbf{\hat{D}_r})_1$. At level two, we note that the three quantities $n \,[\mathbf{\hat{D}_r}; 2]$, $c \,[\mathbf{\hat{D}_r}; 2]$ and $l \,[\mathbf{\hat{D}_r}; 2]$ seem to form three arithmetic progressions (constant difference between consecutive elements)\footnote{This is not fully clear from the table where we have only given two terms in this progression; however we have observed this same progression for $r = 7, 8, 9, 10$ which we have not shown here.}, for $r \geq 5$. We will be able to derive an exact formula for the Wronskian index of all $(\mathbf{\hat{D}_r})_2$ CFTs also which will confirm this observation. From our table we can see that  $(\mathbf{\hat{D}_4})_1$
is a two-character vanishing-Wronskian-index CFT; this indeed appears as one of the solutions in \cite{Mathur:1988na} where the study of two-character vanishing-Wronskian-index MLDE was first made. We can again use the table to anticipate solutions of MLDEs; for example, we can predict two solutions (for two different sets of parameters) for the three-character vanishing-Wronskian-index MLDE viz. the $(\mathbf{\hat{D}_5})_1$ and $(\mathbf{\hat{D}_6})_1$ CFTs.

\subsection{Exceptional series\label{36s}}

\subsubsection{$(\mathbf{\hat{G}_2})_k$ \label{361}}

For the $(\mathbf{\hat{G}_2})_k$ CFT,  the affine algebra has rank three and the finite algebra has rank two; the dimension of the algebra is $14$ and the dual Coxeter number is $4$ (equation \eqref{2g2DG} of appendix \ref{app2}), which when used in \eqref{centralcharge} gives the central charge:
\be \label{44}
c \,[\hat{G}_2; k] = \frac{14\,k}{k + 4}.
\ee
From the Dynkin diagram of $(\mathbf{\hat{G}_2})_k$ (see appendix \ref{app2}), we note that  the co-marks are $2$ and $1$ and hence the integrable representations,  $[\lambda_0, \lambda_1, \lambda_2]$, are restricted by \eqref{level-label}:
\be \label{45}
k = \lambda_0 + 2 \lambda_1 + \lambda_2.
\ee
For every non-negative integer solution to \eqref{29}, there is a primary field in the CFT in the $[\lambda_1, \lambda_2]$ representation. In the absence of any symmetry in the Dynkin diagram, we expect the number of solutions to also be the number $n$ that enters into the formula \eqref{windex}, which we can obtain explicitly:
\bea \label{46}
n \,[\hat{G}_2; k] = & \frac{(k+2)^2}{4}, \qquad \text{for even}~ k \nonumber \\
& \frac{(k+1)(k+3)}{4} \qquad \text{for odd}~ k.
\eea
The conformal dimension of the primary field in the $[\lambda_1, \lambda_2]$ representation, we obtain from \eqref{h}, is given by  
 \be \label{47}
h_{[\lambda_1, \lambda_2]} = \frac{6 \lambda_1^2 + 6 \lambda_1\lambda_2 + 2 \lambda_2^2 + 18 \lambda_1 + 10 \lambda_2}{6 (k + 4)}.
\ee 
If we define
\be \label{48}
D_{[\lambda_1, \lambda_2]} \equiv 6 \lambda_1^2 + 6 \lambda_1\lambda_2 + 2 \lambda_2^2 + 18 \lambda_1 + 10 \lambda_2, \qquad 
D(m) \equiv \sum_{2\lambda_1 + \lambda_2 = m} D_{[\lambda_1, \lambda_2]},
\ee
then the summation over conformal dimensions of the primary fields can be written:
\be \label{49}
\sum_i h_i \equiv \frac{H(k)}{6 (k + 4)}, \qquad H(k) =  \sum_{m=0}^k D(m).
\ee
Now a formula for $D(m)$ can be obtained as
\bea 
D(m) &=& \sum_{n = 0}^{\frac{m}{2}} D_{[\frac{m-2n}{2} , 2n]}  = \frac{m\, (m + 2) (5 m + 29)}{6}~~\qquad \text{for even}~ m \nonumber \\
&=& \sum_{n = 0}^{\frac{m-1}{2}} D_{[\frac{m-2n-1}{2} , 2n+1]}  = \frac{ (m + 2) (10 m^2 + 59 m + 3)}{12}~~\qquad \text{for odd}~ m, \nonumber
\eea
from which one obtains the formula for $H(k)$:
\bea \label{50}
H(k) = & \frac{k (k + 2) (5 k^2 + 49 k + 74)}{24}, \qquad \text{for even}~ k \nonumber \\
& \frac{(k + 1) (k + 3) (k+8) (5 k -1)}{24} \qquad \text{for odd}~ k.
\eea
We can now assemble the equations \eqref{44}, \eqref{46}, \eqref{49}, \eqref{50} to compute the Wronskian index \eqref{windex} of the  $(\mathbf{\hat{G}_2})_k$ CFT:
\bea \label{51}
l \,[\mathbf{\hat{G}_2}; k] = & \frac{k (k^2 - 4) (3 k + 4)}{96}, \qquad \text{for even}~ k \nonumber \\
& \frac{ (k^2 - 1) (k + 3) (3 k - 5)}{96} \qquad \text{for odd}~ k.
\eea
Comparing the $(\mathbf{\hat{G}_2})_k$ CFT results here with the $(\mathbf{\hat{A}_2})_k$ CFT results of section \ref{322}, \eqref{20} and \eqref{51}, we find an unexpected surprise in that 
\be
l \,[\mathbf{\hat{G}_2}; k] = l \,[\mathbf{\hat{A}_2}; k].
\ee
We note this agreement between the two CFTs extends to the number of fields $n \,[\mathbf{\hat{G}_2}; k] = n \,[\mathbf{\hat{A}_2}; k]$,  \eqref{16} and \eqref{46}, which is not hard to understand: a primary field in the $(\mathbf{\hat{A}_2})_k$ CFT in the $[a,b]$ representation (choose $a \leq b$) corresponds in a bijective way to a primary field in the $(\mathbf{\hat{G}_2})_k$ CFT in the $[a, b-a]$ representation. However, what is surprising is that the central charge and the conformal dimensions of the primary fields of the two CFTs are mismatched, but in just the precise way, to make the Wronskian indices agree.

\subsubsection{$(\mathbf{\hat{F}_4})_k$, $(\mathbf{\hat{E}_6})_k$, $(\mathbf{\hat{E}_7})_k$, $(\mathbf{\hat{E}_8})_k$ \label{362}}

For the $(\mathbf{\hat{F}_4})_k$ CFT, the affine algebra has rank five and the finite algebra has rank four; the dimension of the algebra is $52$ and the dual Coxeter number is $9$ (equation \eqref{2f4DG} of appendix \ref{app2}), which when used in \eqref{centralcharge} gives the central charge:
\be \label{52}
c \,[\mathbf{\hat{F}_4}; k] = \frac{52\,k}{k + 9}.
\ee
From the Dynkin diagram of $(\mathbf{\hat{F}_4})_k$ (see appendix \ref{app2}), we note the co-marks and hence the integrable representations,  $[\lambda_0, \lambda_1, \lambda_2, \lambda_3, \lambda_4]$, are restricted by \eqref{level-label}:
\be \label{53}
k = \lambda_0 + 2 \lambda_1 + 3\lambda_2 + 2 \lambda_3 + \lambda_4.
\ee
We use the computer program to obtain the number of primary fields as well as their scaling dimensions and then compute the Wronskian index of the CFT; the results are in table 5. We note that CFTs at levels one and two are the only ones with vanishing Wronskian indices.

For the $(\mathbf{\hat{E}_6})_k$ CFT, the affine algebra has rank seven and the finite algebra has rank six; the dimension of the algebra is $78$ and dual Coxeter number is $12$ (equation \eqref{2e6DG} of appendix \ref{app2}), which when used in \eqref{centralcharge} gives the central charge:
\be \label{54}
c \,[\mathbf{\hat{E}_6}; k] = \frac{78\,k}{k + 12}.
\ee
From the Dynkin diagram of $(\mathbf{\hat{E}_6})_k$ (see appendix \ref{app2}), we note the co-marks and hence the integrable representations,  $[\lambda_0, \lambda_1,  \ldots \lambda_6]$, are restricted by \eqref{level-label}:
\be \label{55}
k = \lambda_0 +  \lambda_1 + 2 \lambda_2 + 3 \lambda_3 + 2 \lambda_4 + \lambda_5 + 2 \lambda_6.
\ee
We note that there is a  symmetry in the Dynkin diagram
\be
\lambda_1 \leftrightarrow  \lambda_5, \qquad \lambda_2 \leftrightarrow  \lambda_4, \nonumber
\ee
which we need to consider when we are counting the number of primary fields. We use the computer program to obtain the number of primary fields as well as their scaling dimensions and then compute the Wronskian index of the CFT; the results are in table 5. We note that CFTs at levels one and two are the only ones with vanishing Wronskian indices.

For the $(\mathbf{\hat{E}_7})_k$ CFT,  the affine algebra has rank eight and the finite algebra has rank seven; the dimension of the algebra is $133$ and dual Coxeter number is $18$ (equation \eqref{2e7DG} of appendix \ref{app2}), which when used in \eqref{centralcharge} gives the central charge:
\be \label{56}
c \,[\mathbf{\hat{E}_7}; k] = \frac{133\,k}{k + 18}.
\ee
From the Dynkin diagram of $(\mathbf{\hat{E}_7})_k$ (see appendix \ref{app2}), we note  the co-marks and hence the integrable representations,  $[\lambda_0, \lambda_1,  \ldots \lambda_7]$, are restricted by \eqref{level-label}:
\be \label{57}
k = \lambda_0 + 2 \lambda_1 + 3 \lambda_2 + 4 \lambda_3 + 3 \lambda_4 + 2\lambda_5 +  \lambda_6 + 2 \lambda_7.
\ee
We use the computer program to obtain the number of primary fields as well as their scaling dimensions and then compute the Wronskian index of the CFT; the results are in table 5. We note that CFTs at levels one and two are the only ones with vanishing Wronskian indices.

For the $(\mathbf{\hat{E}_8})_k$ CFT,  the affine algebra has rank nine and the finite algebra has rank eight; the dimension of the algebra is $248$ and dual Coxeter number is $30$ (equation \eqref{2e8DG} of appendix \ref{app2}), which when used in \eqref{centralcharge} gives the central charge:
\be \label{58}
c \,[\mathbf{\hat{E}_8}; k] = \frac{248\,k}{k + 30}.
\ee
From the Dynkin diagram of $(\mathbf{\hat{E}_8})_k$ (see appendix \ref{app2}), we note  the co-marks and hence the integrable representations,  $[\lambda_0, \lambda_1,  \ldots \lambda_8]$, are restricted by \eqref{level-label}:
\be \label{59}
k = \lambda_0 + 2 \lambda_1 + 3 \lambda_2 + 4 \lambda_3 + 5 \lambda_4 + 6\lambda_5 +  4 \lambda_6 + 2 \lambda_7 + 3 \lambda_8.
\ee
This is the first and only time that we find the absence of a co-mark with value $1$. $\mathbf{E_8}$ is the only simple Lie-algebra for which this happens. This means that at level one, there are no solutions to \eqref{59} except the one in the trivial representation, which has a vanishing conformal dimension. Thus the $(\mathbf{\hat{E}_8})_1$ CFT is a one-character CFT and one-character CFTs necessarily have non-vanishing Wronskian indices. From \eqref{windex}, we can see that the Wronskian index of a one-character CFT is one fourths of the central charge; here the $(\mathbf{\hat{E}_8})_1$ CFT has a central charge of $8$ and hence has a Wronskian index of $2$. 
For higher levels, we use the computer program to obtain the number of primary fields as well as their scaling dimensions and then compute the Wronskian index; the results are in table 5. We note that CFTs at levels two and three have vanishing Wronskian indices. This is the only Lie-algebra for which there is a vanishing-Wronskian-index CFT at level $3$.

\subsection{All Lie algebras at level one \label{37s} \label{371}}

In every computation of a level one CFT, we have found that it has a vanishing Wronskian index. We have noted this fact every time we have encountered it and it can be seen summarily from the first rows of tables 1, 2, 3, 4 and 5. The only exception to this is the $(\mathbf{\hat{E}_8})_1$ CFT which is a one-character CFT.  With these observations, we are led to the surmise that every WZW CFT at level one with two or more characters has a vanishing Wronskian index. We will proceed to show in this section that this surmise is indeed true; we need  only to do this for the classical series. 

In the following, it is useful to introduce the notation $\sigma_i$ for the representation with all Dynkin labels vanishing except the $i$'th Dynkin label, which is $1$. 

\subsubsection{$(\mathbf{\hat{A}_r})_1$}
The central charge for the these CFTs \eqref{21}:
\be \label{60}
c \,[\mathbf{\hat{A}_r}; 1] = r
\ee
and the equation that defines the integrable representations is 
\be \label{61}
1 = \lambda_0 + \lambda_1 + \ldots + \lambda_r.
\ee
The non-negative solutions of \eqref{61} modulo the symmetry \eqref{23} are the trivial representation $[0, \ldots 0]$ and $\sigma_i$, where $i$ takes values from $1$ to $\frac{r}{2}$ when $r$ is even, and from $1$ to $\frac{r+1}{2}$ when $r$ is odd. 
Thus the number of primary fields is given by the formula:
\bea \label{62}
n \,[\mathbf{\hat{A}_r}; 1] = & \frac{r+2}{2} , \qquad \text{for even}~ r \nonumber \\
& \frac{r+3}{2} \qquad \text{for odd}~ r.
\eea
To compute the scaling dimensions of these primary fields, we note the following useful formula:
\be \label{63}
( \sigma_m, \sigma_m + 2\rho)  = F_{mm} + 2 \sum_{i =1}^r F_{mi} 
\ee
where $F_{ij}$ is quadratic form matrix of the Lie-algebra. This formula is true for any Lie-algebra. Here using the explicit form of the quadratic form matrix for $\mathbf{A}_r$, given in the appendix \eqref{2aQ}:
\bea 
( \sigma_m, \sigma_m + 2\rho)  = \frac{m\,(r+1-m)\,(r+2)}{r+1}. \nonumber
\eea
The conformal dimension of the primary field corresponding to the $\sigma_m$ representation is thus, \eqref{h}:
\be \label{64}
h_{\sigma_m} = \frac{m\,(r+1-m)}{2 \,(r+1)}
\ee
Now we can sum over all the conformal dimensions
\bea \label{65}
\sum h &=& \sum_{m = 1}^{\frac{r}{2}} \frac{m\,(r+1-m)}{2 \,(r+1)} = \frac{r (r+2)}{24}, \qquad \qquad \text{for even}~  r \nonumber \\
	   &=& \sum_{m = 1}^{\frac{r+1}{2}} \frac{m\,(r+1-m)}{2 \,(r+1)} = \frac{(r + 3) (2 r + 1)}{48}, \qquad \text{for odd}~  r. 
\eea
We can now assemble the equations \eqref{60}, \eqref{62},  \eqref{65} to compute the Wronskian index \eqref{windex} of the  level one CFTs:
\be \label{66}
l \,[\mathbf{\hat{A}_r}; 1] = 0.
\ee
We have thus shown that for all ranks the $(\mathbf{\hat{A}_r})_1$ CFTs have a vanishing Wronskian index. This expectation, which was based on explicit computations for ranks up to $6$ in table 1, is indeed borne out. 

\subsubsection{$(\mathbf{\hat{B}_r})_1$ \label{372}}
The central charge for the these CFTs \eqref{24}:
\be \label{67}
c \,[\mathbf{\hat{B}_r}; 1] = r + \frac12
\ee
and the equation that defines the integrable representations is 
\be \label{68}
1 = \lambda_0 + \lambda_1 + 2 \lambda_2 + \ldots + 2\lambda_{r-1} + \lambda_r. 
\ee
The non-negative solutions of \eqref{68}  are the trivial representation $[0, \ldots 0]$, $\sigma_1$ and $\sigma_r$. Thus the number of primary fields is given by the formula:
\be \label{69}
n \,[\mathbf{\hat{B}_r}; 1] = 3
\ee
Using \eqref{63} and the first and last rows of the quadratic form matrix for $\mathbf{B}_r$, given in the appendix \eqref{2bQ}, we get  $( \sigma_1, \sigma_1 + 2\rho)  = 2\,r$ and $( \sigma_r, \sigma_r + 2\rho)  = \frac{r \, (2\,r+1)}{4}$ from which the conformal dimensions follow:
\be \label{70}
h_{\sigma_1} = \frac12, \qquad h_{\sigma_r} = \frac{2\,r+1}{16}
\ee
We can now assemble the equations \eqref{67}, \eqref{69},  \eqref{70} to compute the Wronskian index \eqref{windex} of the  level one CFTs:
\be \label{71}
l \,[\mathbf{\hat{B}_r}; 1] = 0.
\ee
We have thus shown that for all ranks the $(\mathbf{\hat{B}_r})_1$ CFTs have a vanishing Wronskian index. This expectation, which was based on explicit computations for ranks up to $6$ in table 2, is indeed borne out.

\subsubsection{$(\mathbf{\hat{C}_r})_1$ \label{373}}
The central charge for the these CFTs \eqref{37}:
\be \label{72}
c \,[\mathbf{\hat{C}_r}; 1] = \frac{r (2 r + 1)}{r+2}
\ee
and the equation that defines the integrable representations is 
\be \label{73}
1 = \lambda_0 + \lambda_1 +  \lambda_2 + \ldots + \lambda_{r-1} + \lambda_r. 
\ee
The non-negative solutions of \eqref{73}  are the trivial representation $[0, \ldots 0]$, $\sigma_1, \sigma_2, \ldots \sigma_r$. Thus the number of primary fields is given by the formula:
\be \label{74}
n \,[\mathbf{\hat{C}_r}; 1] = r + 1.
\ee
We will compute the sum of the conformal dimensions straight away and for that we use \eqref{63} to get 
\be 
\sum_{m = 1} ^{r}( \sigma_m, \sigma_m + 2\rho)  = \text{tr} F + 2 \sum_{i,j} F_{ij}
\ee
and from the explicit form of the quadratic form matrix for $\mathbf{C}_r$, given in the appendix \eqref{2cQ}, we have $\text{tr}~ F = \frac{r\,(r+1)}{4}$ and $2 \sum_{i,j} F_{ij} = \frac{r\,(r + 1)\,(2 r + 1)}{6}$ and finally
\be \label{76}
\sum h = \frac{r\,(r + 1)\,(4 r + 5)}{24\,(r+2)}.
\ee
We can now assemble the equations \eqref{72}, \eqref{74},  \eqref{76} to compute the Wronskian index \eqref{windex} of the  level one CFTs:
\be \label{77}
l \,[\mathbf{\hat{C}_r}; 1] = 0.
\ee
We have thus shown that for all ranks the $(\mathbf{\hat{C}_r})_1$ CFTs have a vanishing Wronskian index. This expectation, which was based on explicit computations for ranks up to $6$ in table 3, is indeed borne out.

\subsubsection{$(\mathbf{\hat{D}_r})_1$ \label{374}}
Among the $(\mathbf{\hat{D}_r})_1$ CFTs, rank four is different from the other ranks because the former has a triality symmetry. A direct computation establishes that the rank four CFT has a vanishing Wronskian index, Here we consider the case $r \geq 5$; the central charge for the these CFTs \eqref{42} is
\be \label{79}
c \,[\mathbf{\hat{D}_r}; 1] = r
\ee
and the equation that defines the integrable representations is 
\be \label{80}
1 = \lambda_0 + \lambda_1 + 2 \lambda_2 + 2 \lambda_3+ \ldots \ldots + 2 \lambda_{r-2} +   \lambda_{r-1} +  \lambda_r.
\ee
The only non-negative solutions to \eqref{80} correspond to the following list of representations for the primary fields: $[0,0, \ldots 0], \quad \sigma_1, \quad \sigma_{r-1} \quad \sigma_r.$ But due to the symmetry in the Dykin diagram of $\mathbf{D}_r$ given in \eqref{43a},  the characters corresponding to the $\sigma_{r-1}$ and $\sigma_r$ representations are expected to be not linearly independent. Thus the number of primary fields is given by the formula:
\be \label{81}
n \,[\mathbf{\hat{D}_r}; 1] = 3.
\ee
Using \eqref{63} and the first and last rows of the quadratic form matrix for $\mathbf{D}_r$, given in the appendix \eqref{2dQ}, we get  $( \sigma_1, \sigma_1 + 2\rho)  = 2\,r-1$ and $( \sigma_r, \sigma_r + 2\rho)  = \frac{r \, (2\,r-1)}{4}$ from which the conformal dimensions follow:
\be \label{82}
h_{\sigma_1} = \frac12, \qquad h_{\sigma_r} = \frac{r}{8}
\ee
We can now assemble the equations \eqref{79}, \eqref{81},  \eqref{82} to compute the Wronskian index \eqref{windex} of the  level one CFTs:
\be \label{83}
l \,[\mathbf{\hat{D}_r}; 1] = 0.
\ee
We have thus shown that for all ranks the $(\mathbf{\hat{D}_r})_1$ CFTs have a vanishing Wronskian index. This expectation, which was based on explicit computations for ranks up to $6$ in table 4, is indeed borne out.

\subsection{All Lie algebras at level two \label{38s}}

In our computations of the Wronskian indices of WZW CFTs done in sections \ref{32s} to \ref{36s} tabulated in tables 1 to 5, we noted some patterns in the results, when we fixed a level and studied all ranks for some classical Lie algebras. One pattern was found in \ref{331} for $(\mathbf{\hat{B}_r})_2$ CFTs where we observed that all the three quantities $n \,[\hat{B}_r; 2]$, $c \,[\hat{B}_r; 2]$ and $l \,[\hat{B}_r; 2]$ formed an arithmetic progression. There is a similar pattern in \ref{352} for $(\mathbf{\hat{D}_r})_2$ CFTs but only when $r \geq 5$. What this means is that the Wronskian indices for these level two CFTs are linear functions of the ranks and we are then motivated to obtain an exact derivation of this linear function. Hence, we study and derive results for all CFTs with classical Lie algebras at level two.

In the following, it is useful to introduce the notation $\tau_i$ for the representation with all Dynkin labels vanishing except the $i$'th Dynkin label, which is $2$.

\subsubsection{$(\mathbf{\hat{A}_r})_2$\label{381}}
The central charge for the these CFTs \eqref{21}:
\be \label{84}
c \,[\mathbf{\hat{A}_r}; 2] = \frac{2\,r(r+2)}{ r + 3}
\ee
and the equation that defines the integrable representations is 
\be \label{85}
2 = \lambda_0 + \lambda_1 + \ldots + \lambda_r.
\ee
The non-negative solutions of \eqref{85} give four classes of representations
\bea \label{86}
\mathbf{I.} \qquad& \text{the trivial representation}\quad [0, \cdots, 0] \nonumber \\
\mathbf{II.} \qquad& \sigma_j, \quad 1 \leq j \leq r \nonumber \\
\mathbf{III.} \qquad& \tau_j, \quad 1 \leq j \leq r \nonumber \\
\mathbf{IV.}  \qquad& \sigma_i + \sigma_j, \quad 1 \leq i < j \leq r.
\eea
We need to count the solutions modulo the symmetry \eqref{23}. To do this, we first identify the representations in \eqref{86} which are fixed points under the action of \eqref{23} and add them to the list of representations in \eqref{86}. This big list contains all the solutions to \eqref{85} modulo the symmetry \eqref{23} but there are exactly  two copies of each solution. To count the number of primary fields, we count the number of representations in this big list and divide by two. The compute the sum of the scaling dimensions of the primary fields, we need to sum the scaling dimensions of all fields in the big list and divide by two. The fixed point representations in \eqref{86} are:
\bea 
\label{87}
\mathbf{V.}\qquad& [0, \cdots, 0], \nonumber  \\
& \sigma_i + \sigma_j,~ i + j = r+1\qquad \text{for even}\quad r \\
\label{88}
\mathbf{V.}\qquad& [0, \cdots, 0], \nonumber  \\
& \sigma_i + \sigma_j,~ i + j = r+1, \nonumber \\
& \sigma_{\frac{r+1}{2}}, \quad \tau_{\frac{r+1}{2}}  \qquad \text{for odd}\quad r.
\eea
The big list is the set of representations $\mathbf{I},\, \mathbf{II},\, \mathbf{III},\,  \mathbf{IV}$ and $\mathbf{V}$. Thus the number of primary fields is given by: 
\bea \label{89}
n \,[\hat{A}_r; 2] = & \frac{r^2 + 4 r + 4}{4}, \qquad \text{for even}~ r \nonumber \\
& \frac{r^2 + 4 r + 7}{4} \qquad \text{for odd}~ r.
\eea
To compute the scaling dimensions of these primary fields, we note the following useful formulae:
\be \label{90}
( \tau_m, \tau_m + 2\rho)  = 4 F_{mm} + 4 \sum_{i =1}^r F_{mi}, \qquad (\sigma_i, \sigma_j) = F_{ij}.
\ee
Using this along with \eqref{63} together with the explicit form the quadratic form matrix for $\mathbf{A}_r$, given in appendix \eqref{2aQ}, we obtain the following formulae for the scaling dimensions \eqref{h} for all fields  in \eqref{86}:
\bea \label{91}
h_{[0, \cdots, 0]} = 0, \quad h_{\sigma_i} = \frac{i(r-i+1)(r+2)}{2(r+1)(r+3)}, \quad  h_{\tau_i} = \frac{i(r-i+1)}{(r+1)} \nonumber \\
h_{\sigma_i + \sigma_j} = \frac{i(r+1)(r+4) + j(r+1)(r+2) - i^2(r+2) - j^2(r+2) - 2 i j}{2(r+1)(r+3)}.
\eea
We can now  compute the sum of the scaling dimensions of the primary fields by  summing the scaling dimensions of all fields in the big list and dividing by two:
\begin{eqnarray} \label{92}
\sum h = &\frac{r(r + 2)(2 r^2 + 11 r + 16)}{48(r+3)} \qquad \text{for even} ~~r\nonumber \\
&\frac{(r + 2)(2 r^3 + 11 r^2 + 22 r + 9)}{48(r+3)} \qquad \text{for odd} ~~r
\end{eqnarray}
We can now assemble the equations \eqref{84}, \eqref{89},  \eqref{92} to compute the Wronskian index \eqref{windex} of the  level two CFTs:
\bea \label{93}
l \,[\hat{A}_r; 2] = & \frac{r(r-2)(r+2)(r+4)}{32}, \qquad \text{for even}~ r \nonumber \\
&  \frac{(r-1)(r+1)^2(r+3)}{32}. \qquad \text{for odd}~ r.
\eea
We note that for even rank, the numerator in the Wronskian index formula is a product of four consecutive even numbers which is always a multiple of $2^4 \times 4! $ and after considering the denominator $32$, the Wronskian index is a multiple of $12$. Similarly for odd rank, the Wronskian index is a multiple of $3$. 

\subsubsection{$(\mathbf{\hat{B}_r})_2$ \label{382}}
The central charge for the these CFTs \eqref{24}:
\be \label{94}
c \,[\mathbf{\hat{B}_r}; 2] = 2 r 
\ee
and the equation that defines the integrable representations is 
\be \label{95}
2 = \lambda_0 + \lambda_1 + 2 \lambda_2 + \ldots + 2\lambda_{r-1} + \lambda_r. 
\ee
The non-negative solutions of \eqref{95} give four classes of representations
\bea \label{96}
\mathbf{I.} \qquad& \text{the trivial representation}\quad [0, \cdots, 0] \nonumber \\
\mathbf{II.} \qquad& \sigma_j, \quad 1 \leq j \leq r \nonumber \\
\mathbf{III.} \qquad& \tau_1, \tau_r  \nonumber \\
\mathbf{IV.}  \qquad& \sigma_1 + \sigma_r. \eea
Thus the number of primary fields is given by the formula:
\be \label{97}
n \,[\mathbf{\hat{B}_r}; 2] = r+4.
\ee
Using \eqref{90} with \eqref{63} together with the explicit form the quadratic form matrix for $\mathbf{B}_r$, given in appendix \eqref{2bQ}, we obtain the following formulae for the scaling dimensions \eqref{h} for all fields  in \eqref{95}:
\bea \label{98}
h_{[0, \cdots, 0]} = 0, \quad h_{\sigma_1} = \frac{r}{2r+1}, \quad h_{\sigma_r} = \frac{r}{8},    \nonumber \\
h_{\sigma_j} = \frac{j}{2} - \frac{j^2}{2(2r+1)}, \quad j = 2, 3, \cdots, r-1 \nonumber \\
h_{\tau_1} = 1, \quad h_{\tau_r} = \frac{r(r+1)}{2(2r+1)} \quad h_{\sigma_1 + \sigma_r} = \frac{r+4}{8}
\eea
We can now  compute the sum of the scaling dimensions of the primary fields:
\begin{eqnarray} \label{99}
\sum h = \frac{2r^2 + 5r + 18}{12}
\end{eqnarray}
We can now assemble the equations \eqref{94}, \eqref{97},  \eqref{99} to compute the Wronskian index \eqref{windex} of the  level two CFTs:
\bea \label{100}
l \,[\mathbf{\hat{B}_r}; 2] = 3 r - 3.
\eea
The equations \eqref{94}, \eqref{97}, \eqref{100}  prove the arithmetic progression that we had observed in table 2.

\subsubsection{$(\mathbf{\hat{C}_r})_2$ \label{383}}
The central charge for the these CFTs \eqref{37}:
\be \label{101}
c \,[\mathbf{\hat{C}_r}; 2] = \frac{2r\,(2 r + 1)}{ r + 3}
\ee
and the equation that defines the integrable representations is 
\be \label{102}
2 = \lambda_0 + \lambda_1 +  \lambda_2 + \ldots  + \lambda_r. 
\ee
The non-negative solutions of \eqref{102} give four classes of representations already given in \eqref{86}; 
the number of primary fields is given by the formula:
\be \label{103}
n \,[\mathbf{\hat{C}_r}; 2] = \frac{(r+2)(r+1)}{2}.
\ee
Using  the explicit form the quadratic form matrix for $\mathbf{C}_r$, given in appendix \eqref{2cQ}, we obtain the following formulae for the scaling dimensions \eqref{h} for all fields:
\bea \label{104}
h_{[0, \cdots, 0]} = 0, \qquad h_{\sigma_j} = \frac{(2r+2)j - j^2}{4(r+3)},    \qquad  h_{\tau_j} = \frac{(4r+6)j - 2j^2}{4(r+3)},   \nonumber \\ h_{\sigma_i + \sigma_j} = \frac{2(r+2)i + 2(r+1)j - i^2 - j^2}{4(r+3)}.
\eea
We can now  compute the sum of the scaling dimensions of the primary fields:
\begin{eqnarray} \label{105}
\sum h = \frac{r (r + 1)(r + 2)(4r + 7)}{24 (r + 3)}
\end{eqnarray}
We can now assemble the equations \eqref{101}, \eqref{103},  \eqref{105} to compute the Wronskian index \eqref{windex} of the  level two CFTs:
\bea \label{106}
l \,[\mathbf{\hat{C}_r}; 2] = \frac{(r-1)r(r+1)(r+2)}{8}.
\eea
We note that the Wronskian index is an integer; in fact it is a multiple of $3$ because the numerator being a product of four consecutive integers is a multiple of $4!$.

\subsubsection{$(\mathbf{\hat{D}_r})_2$ \label{384}}
In the following we will study the $r \geq 5$ theories. The central charge for the these CFTs \eqref{42}:
\be \label{107}
c \,[\mathbf{\hat{D}_r}; 2] = 2 r -1
\ee
and the equation that defines the integrable representations is 
\be \label{108}
2 = \lambda_0 + \lambda_1 + 2 \lambda_2 + 2 \lambda_3+ \ldots \ldots + 2 \lambda_{r-2} +   \lambda_{r-1} +  \lambda_r.
\ee
The non-negative solutions of \eqref{108}, after considering the symmetry \eqref{43a}, give four classes of representations
\bea \label{109}
\mathbf{I.} \qquad& \text{the trivial representation}\quad [0, \cdots, 0] \nonumber \\
\mathbf{II.} \qquad& \sigma_j, \quad 1 \leq j \leq r-1 \nonumber \\
\mathbf{III.} \qquad& \tau_1, \tau_{r-1}  \nonumber \\
\mathbf{IV.}  \qquad& \sigma_1 + \sigma_{r-1}, ~\sigma_{r-1} + \sigma_{r}. \eea
Thus the number of primary fields is given by the formula:
\be \label{110}
n \,[\mathbf{\hat{D}_r}; 2] = r+4.
\ee
Using \eqref{90} with \eqref{63} together with the explicit form the quadratic form matrix for $\mathbf{D}_r$, given in appendix \eqref{2dQ}, we obtain the following formulae for the scaling dimensions \eqref{h} for all fields  in \eqref{109}:
\bea \label{111}
h_{[0, \cdots, 0]} = 0, \quad h_{\sigma_1} = \frac{2r-1}{4r}, \quad h_{\sigma_{r-1}} = \frac{2r-1}{16},    \nonumber \\
h_{\sigma_j} = \frac{j}{2} - \frac{j^2}{4r}, \quad j = 2, 3, \cdots, r-2 \nonumber \\
h_{\tau_1} = 1, \quad h_{\tau_{r-1}} = \frac{r}{4}, \quad h_{\sigma_1 + \sigma_{r-1}} = \frac{2r+7}{16}, \quad \quad h_{\sigma_{r-1} + \sigma_{r}} = \frac{r^2 - 1}{4r}.
\eea
The sum of the scaling dimensions of the primary fields:
\begin{eqnarray} \label{112}
\sum h = \frac{4r^2 + 9r + 32}{24}
\end{eqnarray}
We can now assemble the equations \eqref{107}, \eqref{110},  \eqref{112} to compute the Wronskian index \eqref{windex} of the  level two CFTs:
\bea \label{113}
l \,[\mathbf{\hat{D}_r}; 2] = 3 r - 3, \qquad \qquad r \geq 5
\eea
The equations \eqref{107}, \eqref{110}, \eqref{113}  prove the arithmetic progression that we had observed in table 4.

\subsection{Summary of results for WZW CFTs\label{39s}}

Here, we will gather all the results on WZW CFTs that we have obtained so far; some of the following are just observations that are yet to be proved \cite{wp1}.

\textbf{1.} We have obtained \emph{exact formulae} for the $n$ and $l$ values of WZW CFTs based on \emph{rank one} Lie algebras at \emph{all levels} viz.  $(\mathbf{\hat{A}_1})_k$ in \ref{321}; all of these CFTs have a vanishing Wronskian-index. 

\textbf{2.} We have obtained \emph{exact formulae} for the $n$ and $l$ values of WZW CFTs based on \emph{rank two} Lie algebras at \emph{all levels} viz.  $(\mathbf{\hat{A}_2})_k$, $(\mathbf{\hat{C}_2})_k$ and $(\mathbf{\hat{G}_2})_k$ in \ref{322}, \ref{341} and \ref{361} respectively. 

\textbf{3.} We have found a remarkable coincidence between the $(n, l)$ values for the $(\mathbf{\hat{A}_2})_k$ and $(\mathbf{\hat{G}_2})_k$ CFTs:  \emph{for every level, their $(n, l)$ values match}.  What this means is they are solutions to the same MLDE (for different sets of parameters in the MLDE). We can thus predict that if for some MLDE, if $(\mathbf{\hat{A}_2})_k$ is found to be a solution for some $k$, then $(\mathbf{\hat{G}_2})_k$ will also be a solution. We can see a special case of this in the solutions of the two-character vanishing-Wronskian-index MLDE of \cite{Mathur:1988na}: both $(\mathbf{\hat{A}_2})_1$ and $(\mathbf{\hat{G}_2})_1$ are there. See also \eqref{145}.

\textbf{4.} We have computed, with the aid of a computer program, the $(n, l)$ values of WZW CFTs with all simple Lie algebras (upto rank $6$ for the classical algebras) and upto level 12 (for all the simple Lie algebras) and these are tabulated in tables 1 - 5.

\textbf{5.} We find another remarkable coincidence between the $(n,l)$ values for the $(\mathbf{\hat{B}_r})_k$ and $(\mathbf{\hat{D}_r})_k$ CFTs, but for $r \geq 5$. This can be seen for $r = 5$ and $r=6$ in tables 2 and 4. The agreement between the number of characters is not hard to explain: a WZW primary field in the $(\mathbf{\hat{B}_r})_k$ CFT in the $[a_1, a_2, \cdots, a_{r-1}, a_r]$ representation  corresponds in a bijective way to  a WZW primary field in the $(\mathbf{\hat{D}_r})_k$ CFT in the $[a_1, a_2, \cdots, a_{r-1}, a_{r-1} + a_r]$ representation. The central charge and the conformal dimensions of the two CFTs are mismatched but just in the precise way necessary to make the Wronskian indices agree. We hope to show and explain this remarkable fact by analytical computation \cite{wp1}.

\textbf{6.} We have obtained \emph{exact formulae} for the $n$ and $l$ values of WZW CFTs based on \emph{all} Lie algebras at \emph{level one} in \ref{37s}; all of these CFTs have a vanishing Wronskian-index. 

\textbf{7.} We have obtained \emph{exact formulae} for the $n$ and $l$ values of WZW CFTs based on \emph{all} Lie algebras at \emph{level two} in \ref{38s}.

\section{Minimal Model CFTs \label{4s}}

\subsection{Virasoro minimal models\label{41s}}

The Virasoro minimal models are a series of rational conformal field theories which are labelled by two co-prime positive integers  and are denoted by $\mathcal{M} (p, p')$ with $p > p' > 1$. These models include unitary and non-unitary CFTs; the choice $p = p'+1$ gives unitary CFTs and all other choices give non-unitary CFTs.  The central charge of the  $\mathcal{M} (p, p')$ CFT is:
\be \label{114}
c \, [\mathcal{M} (p, p')] = 1 -  \frac{6\,(p - p^\prime)^2}{p\, p^\prime}.
\ee
These CFTs have Virasoro primary fields $\mathcal{O}_{r,s}$ whose scaling dimensions are denoted $h_{r,s}$: \be \label{116}
h_{r, s} = \frac{(p r - p^\prime s)^2 - (p - p^\prime)^2}{4p p^\prime}, \qquad 1\leq r \leq p^\prime-1,\quad 1\leq s \leq p-1.
\ee
This formula for the conformal dimensions has the following symmetry,
\be
h_{r, s} = h_{p^\prime - r, p - s},
\ee
(which does not have any fixed points since  a fixed point  requires both $p$ and $p^\prime$ to be even); there is an identification $\mathcal{O}_{r, s} = \mathcal{O}_{p^\prime - r, p - s}$ to be made and hence the number of primary fields is given by:
\be \label{115}
n \, [\mathcal{M} (p, p')] = \frac{(p-1)(p^\prime - 1)}{2}.
\ee
To obtain the sum of the conformal dimensions, we sum over the whole range of $r$ and $s$ in \eqref{116} and  divide the answer by two:
\be \label{118}
\sum h = \frac12 \sum\limits_{r=1}^{(p^\prime - 1)}\sum\limits_{s=1}^{(p-1)} h_{r, s} = \frac{(p-1)(p^\prime - 1)(pp^\prime - p - p^\prime)}{48} - \frac{(p-p^\prime)^2(p-1)(p^\prime-1)}{8pp^\prime} 
\ee
It is remarkable that when we assemble \eqref{114}, \eqref{115} and \eqref{118} into the Wronskian index formula \eqref{windex}, it vanishes for all $\mathcal{M} (p, p')$ CFTs.
\be
l \, [\mathcal{M} (p, p')] = 0.
\ee
This series of CFTs includes a one-character CFT viz. $\mathcal{M} (3, 2)$ which turns out to be a trivial CFT with vanishing central charge. The Virasoro series has a single two-character CFT viz. $\mathcal{M} (5, 2)$ which is a non-unitary CFT with central charge $-\frac{22}{5}$. This two-character vanishing-Wronskian-index CFT should have shown up as a solution to the two-character vanishing-Wronskian-index MLDE and indeed it does \cite{Mathur:1988na}. The Virasoro series has two three-character CFTs viz. the non-unitary $\mathcal{M} (7, 2)$ and the unitary $\mathcal{M} (4, 3)$ and these should provide two solutions (for different values of parameters) to the three-character vanishing-Wronskian-index MLDE. Similarly there are two four-character CFTs, two five-character CFTs, three six-character CFTs \ldots. all of which are solutions to the corresponding vanishing-Wronskian index MLDEs.

\subsection{$\mathcal{N} = 1$ Super-Virasoro minimal models \label{42s}}
The $\mathcal{N} = 1$ super-Virasoro minimal models are a series of unitary rational conformal field theories which are labelled by a positive integer $m$ which takes values  $3, 4, 5, \cdots$ and denoted by $\mathcal{SM}(m+2,m)$ \cite{Friedan:1983xq}. These theories have the $\mathcal{N} = 1$ super  conformal symmetry which is a super Lie algebra containing the usual Virasoro modes together with super current modes (both Ramond and Neveu-Schwarz).
The central charge of the  $\mathcal{SM} (m+2, m)$ CFT is:
\be \label{120}
c \, [\mathcal{SM} (m+2, m)] = \frac32 -  \frac{12}{m (m+2)}.
\ee
The  super conformal primary fields $\mathcal{O}_{r,s}$ of these CFTs have conformal dimensions:
\be \label{121}
h_{r, s} = \frac{((m+2)\,r -  m\,s)^2 - 4}{8m (m+2)} + \frac{\epsilon}{16}, \qquad 1\leq r < m, \quad 1\leq s < m+2.
\ee
When $r - s$ is even, $\mathcal{O}_{r,s}$ is a Neveu-Schwarz field and one has to set $\epsilon = 0$ and when $r - s$ is odd, $\mathcal{O}_{r,s}$ is a Ramond field and one has to set $\epsilon = 1$ above. The formula for the conformal dimensions has the following symmetry: 
\be \label{122}
h_{r, s} = h_{m - r, m + 2 - s},
\ee
and there is an identification $\mathcal{O}_{r, s} = \mathcal{O}_{m - r, m + 2 - s}$ to be made. To obtain the number of super conformal primaries we add to the set of fields $\mathcal{O}_{r,s}$ ($r$ and $s$ taking values as in \eqref{121}) the set of fixed points under the \eqref{122} symmetry and dividing the total number of fields by two. When $m$ is odd, there are no fixed points and when $m$ is even, there is one fixed point $h_{\frac{m}{2}, \frac{m+2}{2}}$. Let $N_{scp}$ denote the number of super conformal primaries; we have 
\bea \label{123}
N_{scp}\,[\mathcal{SM} (m+2, m)] ~=~ &\frac{m^2-1}{2} \qquad \text{for odd} \quad m \nonumber \\
 &\frac{m^2}{2} \qquad \text{for even} \quad m.
\eea
In each case, the number of super conformal primaries is an even number; half of them are Neveu-Schwarz  fields and the other half  are Ramond fields.  Now, to compute how many conformal primary fields there are in the $\mathcal{SM} (m+2, m)$ CFT, we need to know how a super conformal representation decomposes into conformal representations \cite{Friedan:1984rv}, \cite{Yang:1987bj}, \cite{Dijkgraaf:1989hb}, \cite{Minces:1998vc}. First we note how Neveu-Schwarz super conformal representations decompose:
\be \label{124}
\left[0\right]_{\rm NS} = \left[0\right]_{\rm V} \oplus \left[\frac{3}{2}\right]_{\rm V}, \quad 
\left[h\right]_{\rm NS} = \left[h\right]_{\rm V} \oplus \left[h + \frac{1}{2}\right]_{\rm V}, \quad h\neq 0.
\ee
A Neveu-Schwarz super conformal primary field of vanishing conformal dimension (such as $h_{1,1}$) splits into two conformal primary fields with conformal dimensions $0$ and $\frac32$. A Neveu-Schwarz super conformal primary field of non-vanishing conformal dimension, say $h$,  splits into two conformal primary fields with conformal dimensions $h$ and $h + \frac12$. Then we note how Ramond super conformal representations decompose: 
\be \label{125}
\left[h\right]_{\rm R} = \left[h\right]_{\rm V}   \oplus \left[h + 1\right]_{\rm V}, ~\text{for} ~h = \frac{c}{24}, \qquad \qquad 
\left[h\right]_{\rm R} = \left[h\right]_{\rm V}  , ~\text{for}~h \neq \frac{c}{24} 
\ee
Every Ramond super conformal primary field decomposes into just one conformal primary field except the one whose conformal dimension is $\frac{c}{24}$ which splits into two conformal primary fields. We note that this happens in $\mathcal{SM} (m+2, m)$ CFTs for even $m$: $h_{\frac{m}{2}, \frac{m+2}{2}}$ computes to $\frac{c}{24}$  \eqref{121}. We can now count the number of primary fields by  using the decomposition rules \eqref{124} and \eqref{125}:
\bea \label{126}
n\,[\mathcal{SM} (m+2, m)] ~=~ &\frac{3}{4} (m^2 - 1) \qquad \text{for odd} \quad m \nonumber \\
 &\frac{3m^2}{4}   + 1 \qquad \text{for even} \quad m.
\eea
We then compute the sum of the conformal dimensions, first we consider the primary fields coming from the Neveu-Schwarz fields. Using \eqref{124}, this is 
\be \label{127}
\sum h \bigg|_{\text{Neveu-Schwarz}} = \underbrace{\sum\limits_r^{m-1}\sum\limits_s^{m+1}}_\text{$(r-s)$ is even} h_{r, s}\, + \frac{N_{scp}-1}{2} + \frac{3}{2}.
\ee
For the primary fields coming from the Ramond fields, the even and odd $m$s are different; for even $m$ we have one of the Ramond fields having a scaling dimension equal to $\frac{c}{24}$ and hence resulting in two primary fields,
\bea \label{128}
\sum h \bigg|_{\text{Ramond}} = &&\frac{1}{2}\left(\underbrace{\sum\limits_r^{m-1}\sum\limits_s^{m+1}}_\text{$(r-s)$ is odd} h_{r, s}\right) \qquad \qquad \text{for odd}~ m \nonumber \\
  && \frac{1}{2}\left(\underbrace{\sum\limits_r^{m-1}\sum\limits_s^{m+1}}_\text{$(r-s)$ is odd} h_{r, s} - h_{\frac{m}{2}, \frac{m+2}{2}}\right) + 2h_{\frac{m}{2}, \frac{m+2}{2}} + 1, \quad \text{for even} ~m.
\eea
We can now assemble the equations \eqref{120}, \eqref{126},  \eqref{127}, \eqref{128} to compute the Wronskian index \eqref{windex}:
\bea \label{100}
l \,[\mathcal{SM}(m+2,m)] = &\frac{ 3(m^2 - 9)(m^2 + 3)}{16} \qquad \qquad \text{for odd}~m \nonumber \\
 & \frac{3(m^2   -  8)(m^2 + 8)}{16} \qquad \qquad \text{for even}~m.
\eea
The Wronskian index thus computed is guaranteed to be an integer since each of the non-constant factors in the numerator (in both even and odd cases) is a multiple of $4$. We note that only $\mathcal{SM}(5,3)$ CFT has a vanishing Wronskian index. This is a six-character CFT which is identical to the Virasoro minimal model $\mathcal{M}(5,4)$. All other $\mathcal{SM}(m+2,m)$ CFTs i.e $m \neq 3$ have a non-vanishing Wronskian index.

\section{Conclusion and Future Directions \label{5s}}

In this paper, we have computed the number of characters and the Wronskian index, i.e. the $(n, l)$ values for a host of known RCFTs viz. the WZW CFTs in section \ref{3s}, the Virasoro minimal model CFTs in \ref{41s} and the $\mathcal{N} = 1$ super-Virasoro CFTs in \ref{42s}. 

We can revisit some of the motivating questions for this work that we gave in section \ref{1s}. First let us discuss in the context of $(n, l) = (2, 0)$ i.e. what are all the known RCFTs that will be solutions to a two-character vanishing vanishing Wronskian-index MLDE. We can just go through results and produce the list: two $\mathbf{A}$-series CFTs viz. $(\mathbf{\hat{A}_1})_1$, $(\mathbf{\hat{A}_2})_1$, one $\mathbf{D}$-series CFT viz. $(\mathbf{\hat{D}_4})_1$,  four  exceptional series CFTs viz. $(\mathbf{\hat{G}_2})_1$, $(\mathbf{\hat{F}_4})_1$, $(\mathbf{\hat{E}_6})_1$, $(\mathbf{\hat{E}_7})_1$,  and one Virasoro minimal model viz. $\mathcal{M}(5,2)$. Even without solving the MLDE, we can be sure that these eight will be solutions. But of course, what \cite{Mathur:1988na} achieves is a classification, the statement that these are the only two-character vanishing-Wronskian-index CFTs.

We can repeat the exercise of the previous paragraph for $(n, l) = (3, 0)$ i.e for three-character vanishing-Wronskian-index RCFTs, where the final classification has not yet been achieved, even if lots of progress has happened \cite{Mathur:1988gt, Mukhi:2020gnj, Tener:2016lcn, kaneko4}. From our results, we obtain the following partial classification of three-character vanishing-Wronskian-index CFTs:
\begin{gather} \label{144}
(\mathbf{\hat{A}_1})_2,\qquad (\mathbf{\hat{A}_3})_1,\qquad (\mathbf{\hat{A}_4})_1, \nonumber \\
(\mathbf{\hat{B}_r})_1 \quad \text{for all} \quad r \geq 3, \nonumber \\
(\mathbf{\hat{C}_2})_1, \nonumber \\
(\mathbf{\hat{D}_r})_1 \quad \text{for all} \quad r \geq 4, \nonumber \\
(\mathbf{\hat{E}_8})_2, \nonumber \\
\mathcal{M}(7,2), \qquad \mathcal{M}(4,3), \nonumber \\
\mathcal{M}(5,2)\otimes\mathcal{M}(5,2), \qquad (\mathbf{\hat{A}_1})_1 \otimes (\mathbf{\hat{A}_1})_1,\qquad (\mathbf{\hat{A}_2})_1 \otimes (\mathbf{\hat{A}_2})_1, \qquad (\mathbf{\hat{D}_4})_1 \otimes (\mathbf{\hat{D}_4})_1, \nonumber \\ 
(\mathbf{\hat{E}_6})_1 \otimes (\mathbf{\hat{E}_6})_1,\qquad (\mathbf{\hat{E}_7})_1 \otimes (\mathbf{\hat{E}_7})_1, \qquad (\mathbf{\hat{F}_4})_1 \otimes (\mathbf{\hat{F}_4})_1,\qquad (\mathbf{\hat{G}_2})_1 \otimes (\mathbf{\hat{G}_2})_1.
\end{gather}

We thus have two infinite series of CFTs and fifteen\footnote{The last eight CFTs are tensor product CFTs. Each of the $(2,0)$ CFTs when tensored with itself gives a CFT with three characters and vanishing Wronskian index (using the Hampapura-Mukhi tensor product formulae from \cite{Hampapura:2015cea}).} ``discrete'' CFTs in this (partial) classification. We note that there are an infinite number of three-character vanishing-Wronskian-index CFTs with Kac-Moody symmetry. 

Continuing and this time considering the $(n, l) = (2, 2)$ case, we find from our results that there is not a single such CFT; we also considered tensor products of known CFTs and still there is not a single CFT. What this means is that if one can find solutions to two-character Wronskian-index-equalling-two MLDE, one would have discovered new CFTs. This is what was achieved in \cite{Naculich:1988xv, Hampapura:2015cea, Gaberdiel:2016zke, Hampapura:2016mmz}.

For the $(n, l) = (4, 0)$ case, the partial classification that we can obtain from our results, consists of only eight CFTs:
\be \label{145}
(\mathbf{\hat{A}_1})_3 \qquad (\mathbf{\hat{A}_2})_2 \qquad (\mathbf{\hat{A}_6})_1 \qquad (\mathbf{\hat{A}_5})_1\qquad (\mathbf{\hat{G}_2})_2 \qquad (\mathbf{\hat{C}_3})_1 \qquad \mathcal{M}(9,2) \qquad \mathcal{M}(5,3)
\ee
We thus see that a partial classification for RCFTs can be achieved with the results of this paper. 

In this paper, we have studied only three classes of RCFTs. There are clearly many more. We have not considered $\mathcal{N} = 2$ superconformal minimal models. Even in an early paper like \cite{Anderson:1987ge} there are six classes of RCFTs listed; we have considered only two of them here in this paper. Studying them along the lines of this paper will expand the partial classifications of this paper. We hope to take up these studies \cite{wp1}.

Given values for $(n , l)$, the MLDE gets fixed except for a few parameters. Solving the MLDE means answering the question: for what values of these parameters does the MLDE have character-like solutions? It may be useful to know the values of the parameters in the MLDE corresponding to the solutions in our partial classification i.e corresponding to the known RCFTs. This knowledge of the positions of known RCFTs in the space of parameters of the MLDE may reveal some structure and offer hints as to where unknown RCFTs could be found or not found. This is a future direction of research based on this work.

With regard to WZW CFTs, our work is incomplete in many directions. We have indicated some of them in \ref{39s}. We have obtained exact formulae, in the classical case,  only for either small ranks (up to $2$) and all levels or for all ranks and small levels  (up to $2$). We hope it is possible to remedy this situation and produce exact formulae for all ranks and all levels. 

\begin{center}
\textbf{Acknowledgments}
\end{center}
We thank Prof. Sunil Mukhi from whose lectures, given in NISER, Bhubaneswar,  and elsewhere, we learnt this subject and also for discussions and answering our queries.  AD would like to thank School of Physics, NISER, Bhubaneswar for the hospitality and resources required to complete this project. AD would also like to thank A. R. Chandra for useful discussions over emails.

\appendix

\section{Results} \label{app1}

Here we tabulate the results of the computations for the $\mathbf{A}$-series in table 1, $\mathbf{B}$-series in table 2, $\mathbf{C}$-series in table 3, $\mathbf{D}$-series in table 4 and the exceptional Lie algebras in table 5. For the classical series we give results up to rank $6$ and in all cases, we give results up to level $12$.

\begin{table}[h] 
\begin{center}
\begin{threeparttable}
\caption{A series}
\begin{tabular*}{\textwidth}{@{\extracolsep{\fill}}cccccccccc}
\hline
\hline
\multicolumn{1}{c}{} & \multicolumn{3}{c}{$A_2$} & \multicolumn{3}{c}{$A_3$} & \multicolumn{3}{c}{$A_4$} \\
\cline{2-4} \cline{5-7} \cline{8-10}
\multicolumn{1}{c}{k} & \multicolumn{1}{c}{n} & \multicolumn{1}{c}{c} & \multicolumn{1}{c}{l} & \multicolumn{1}{c}{n} & \multicolumn{1}{c}{c} & \multicolumn{1}{c}{l} & \multicolumn{1}{c}{n} & \multicolumn{1}{c}{c} & \multicolumn{1}{c}{l} \\ 
\hline
\makebox[0pt][l]{\fboxsep0pt\colorbox{Mywhite} {\strut\hspace*{\linewidth}}}
1 & 2 & 2 & 0 & 3 & 3 & 0 & 3 & 4 & 0  \\
\makebox[0pt][l]{\fboxsep0pt\colorbox{Mygrey} {\strut\hspace*{\linewidth}}}
2 & 4 & 16/5 & 0 & 7 & 5 & 6 & 9 & 48/7 & 12  \\
\makebox[0pt][l]{\fboxsep0pt\colorbox{Mywhite} {\strut\hspace*{\linewidth}}}
3 & 6 & 4 & 2 & 13 & 45/7 & 36 & 19 & 9 & 96  \\
\makebox[0pt][l]{\fboxsep0pt\colorbox{Mygrey} {\strut\hspace*{\linewidth}}}
4 & 9 & 32/7 & 8 & 22 & 15/2 & 135 & 38 & 32/3 & 495  \\
\makebox[0pt][l]{\fboxsep0pt\colorbox{Mywhite} {\strut\hspace*{\linewidth}}}
5 & 12 & 5 & 20 & 34 & 25/3 & 375 & 66 & 12 & 1695  \\
\makebox[0pt][l]{\fboxsep0pt\colorbox{Mygrey} {\strut\hspace*{\linewidth}}}
6 & 16 & 16/3 & 44 & 50 & 9 & 895 & 110 & 144/11 & 5083  \\
\makebox[0pt][l]{\fboxsep0pt\colorbox{Mywhite} {\strut\hspace*{\linewidth}}}
7 & 20 & 28/5 & 80 & 70 & 105/11 & 1875 & 170 & 14 & 12723  \\
\makebox[0pt][l]{\fboxsep0pt\colorbox{Mygrey} {\strut\hspace*{\linewidth}}}
8 & 25 & 64/11 & 140 & 95 & 10 & 3625 & 255 & 192/13 & 29553  \\
\makebox[0pt][l]{\fboxsep0pt\colorbox{Mywhite} {\strut\hspace*{\linewidth}}}
9 & 30 & 6 & 220 & 125 & 135/13 & 6505 & 365 & 108/7 & 61873  \\
\makebox[0pt][l]{\fboxsep0pt\colorbox{Mygrey} {\strut\hspace*{\linewidth}}}
10 & 36 & 80/13 & 340 & 161 & 75/7 & 11095 & 511 & 16 & 123193  \\
\makebox[0pt][l]{\fboxsep0pt\colorbox{Mywhite} {\strut\hspace*{\linewidth}}}
11 & 42 & 44/7 & 490 & 203 & 11 & 18025 & 693 & 33/2 & 229173  \\
\makebox[0pt][l]{\fboxsep0pt\colorbox{Mygrey} {\strut\hspace*{\linewidth}}}
12 & 49 & 32/5 & 700 & 252 & 45/4 & 28266 & 924 & 288/17 & 410970  \\
\hline
\hline
\end{tabular*} 
\end{threeparttable}
\end{center}
\end{table}

\begin{table}[h]
\begin{center}
\begin{threeparttable}
\begin{tabular*}{\textwidth}{@{\extracolsep{\fill}}ccccccc}
\hline
\hline
\multicolumn{1}{c}{} & \multicolumn{3}{c}{$A_5$} & \multicolumn{3}{c}{$A_6$} \\
\cline{2-4} \cline{5-7}
\multicolumn{1}{c}{k} & \multicolumn{1}{c}{n} & \multicolumn{1}{c}{c} & \multicolumn{1}{c}{l} & \multicolumn{1}{c}{n} & \multicolumn{1}{c}{c} & \multicolumn{1}{c}{l} \\ 
\hline
\makebox[0pt][l]{\fboxsep0pt\colorbox{Mywhite} {\strut\hspace*{\linewidth}}}
1 & 4 & 5 & 0 & 4 & 6 & 0  \\
\makebox[0pt][l]{\fboxsep0pt\colorbox{Mygrey} {\strut\hspace*{\linewidth}}}
2 & 13 & 35/4 & 36 & 16 & 32/3 & 60  \\
\makebox[0pt][l]{\fboxsep0pt\colorbox{Mywhite} {\strut\hspace*{\linewidth}}}
3 & 32 & 35/3 & 340 & 44 & 72/5 & 700  \\
\makebox[0pt][l]{\fboxsep0pt\colorbox{Mygrey} {\strut\hspace*{\linewidth}}}
4 & 70 & 14 & 1955 & 110 & 192/11 & 5155  \\
\makebox[0pt][l]{\fboxsep0pt\colorbox{Mywhite} {\strut\hspace*{\linewidth}}}
5 & 136 & 175/11 & 8060 & 236 & 20 & 25480  \\
\makebox[0pt][l]{\fboxsep0pt\colorbox{Mygrey} {\strut\hspace*{\linewidth}}}
6 & 246 & 35/2 & 27695 & 472 & 288/13 & 105720  \\
\makebox[0pt][l]{\fboxsep0pt\colorbox{Mywhite} {\strut\hspace*{\linewidth}}}
7 & 416 & 245/13 & 81500 & 868 & 24 & 364620  \\
\makebox[0pt][l]{\fboxsep0pt\colorbox{Mygrey} {\strut\hspace*{\linewidth}}}
8 & 671 & 20 & 215885 & 1519 & 128/5 & 1129545  \\
\makebox[0pt][l]{\fboxsep0pt\colorbox{Mywhite} {\strut\hspace*{\linewidth}}}
9 & 1036 & 21 & 520660 & 2520 & 27 & 3130320  \\
\makebox[0pt][l]{\fboxsep0pt\colorbox{Mygrey} {\strut\hspace*{\linewidth}}}
10 & 1547 & 175/8 & 1170141 & 4032 & 480/17 & 8048856  \\
\makebox[0pt][l]{\fboxsep0pt\colorbox{Mywhite} {\strut\hspace*{\linewidth}}}
11 & 2240 & 385/17 & 2466744 & 6216 & 88/3 & 19184568  \\
\makebox[0pt][l]{\fboxsep0pt\colorbox{Mygrey} {\strut\hspace*{\linewidth}}}
12 & 3164 & 70/3 & 4940754 & 9324 & 576/19 & 43248258  \\
\hline
\hline
\end{tabular*} 
\end{threeparttable}
\end{center}
\end{table}

\begin{table}[h]
\begin{center}
\begin{threeparttable}
\caption{B series}
\begin{tabular*}{\textwidth}{@{\extracolsep{\fill}}ccccccc}
\hline
\hline
\multicolumn{1}{c}{} & \multicolumn{3}{c}{$B_3$} & \multicolumn{3}{c}{$B_4$} \\
\cline{2-4} \cline{5-7}
\multicolumn{1}{c}{k} & \multicolumn{1}{c}{n} & \multicolumn{1}{c}{c} & \multicolumn{1}{c}{l} & \multicolumn{1}{c}{n} & \multicolumn{1}{c}{c} & \multicolumn{1}{c}{l} \\ 
\hline
\makebox[0pt][l]{\fboxsep0pt\colorbox{Mywhite} {\strut\hspace*{\linewidth}}}
1 & 3 & 7/2 & 0 & 3 & 9/2 & 0 \\
\makebox[0pt][l]{\fboxsep0pt\colorbox{Mygrey} {\strut\hspace*{\linewidth}}}
2 & 7 & 6 & 6 & 8 & 8 & 9 \\
\makebox[0pt][l]{\fboxsep0pt\colorbox{Mywhite} {\strut\hspace*{\linewidth}}}
3 & 13 & 63/8 & 36 & 16 & 54/5 & 63 \\
\makebox[0pt][l]{\fboxsep0pt\colorbox{Mygrey} {\strut\hspace*{\linewidth}}}
4 & 22 & 28/3 & 135 & 30 & 144/11 & 288 \\
\makebox[0pt][l]{\fboxsep0pt\colorbox{Mywhite} {\strut\hspace*{\linewidth}}}
5 & 34 & 21/2 & 375 & 50 & 15 & 918 \\
\makebox[0pt][l]{\fboxsep0pt\colorbox{Mygrey} {\strut\hspace*{\linewidth}}}
6 & 50 & 126/11 & 895 & 80 & 216/13 & 2563 \\
\makebox[0pt][l]{\fboxsep0pt\colorbox{Mywhite} {\strut\hspace*{\linewidth}}}
7 & 70 & 49/4 & 1875 & 120 & 18 & 6093 \\
\makebox[0pt][l]{\fboxsep0pt\colorbox{Mygrey} {\strut\hspace*{\linewidth}}}
8 & 95 & 168/13 & 3625 & 175 & 96/5 & 13468 \\
\makebox[0pt][l]{\fboxsep0pt\colorbox{Mywhite} {\strut\hspace*{\linewidth}}}
9 & 125 & 27/2 & 6505 & 245 & 81/4 & 27118 \\
\makebox[0pt][l]{\fboxsep0pt\colorbox{Mygrey} {\strut\hspace*{\linewidth}}}
10 & 161 & 14 & 11095 & 336 & 360/17 & 52038 \\
\makebox[0pt][l]{\fboxsep0pt\colorbox{Mywhite} {\strut\hspace*{\linewidth}}}
11 & 203 & 231/16 & 18025 & 448 & 22 & 93898 \\
\makebox[0pt][l]{\fboxsep0pt\colorbox{Mygrey} {\strut\hspace*{\linewidth}}}
12 & 252 & 252/17 & 28266 & 588 & 432/19 & 163632 \\
\hline
\hline
\end{tabular*} 
\end{threeparttable}
\end{center}
\end{table}

\begin{table}[h]
\begin{center}
\begin{threeparttable}
\begin{tabular*}{\textwidth}{@{\extracolsep{\fill}}ccccccc}
\hline
\hline
\multicolumn{1}{c}{} & \multicolumn{3}{c}{$B_5$} & \multicolumn{3}{c}{$B_6$} \\
\cline{2-4} \cline{5-7}
\multicolumn{1}{c}{k} & \multicolumn{1}{c}{n} & \multicolumn{1}{c}{c} & \multicolumn{1}{c}{l} & \multicolumn{1}{c}{n} & \multicolumn{1}{c}{c} & \multicolumn{1}{c}{l} \\ 
\hline
\makebox[0pt][l]{\fboxsep0pt\colorbox{Mywhite} {\strut\hspace*{\linewidth}}}
1 & 3 & 11/2 & 0 & 3 & 13/2 & 0 \\
\makebox[0pt][l]{\fboxsep0pt\colorbox{Mygrey} {\strut\hspace*{\linewidth}}}
2 & 9 & 10 & 12 & 10 & 12 & 15 \\
\makebox[0pt][l]{\fboxsep0pt\colorbox{Mywhite} {\strut\hspace*{\linewidth}}}
3 & 19 & 55/4 & 96 & 22 & 117/7 & 135 \\
\makebox[0pt][l]{\fboxsep0pt\colorbox{Mygrey} {\strut\hspace*{\linewidth}}}
4 & 39 & 220/13 & 525 & 49 & 104/5 & 870 \\
\makebox[0pt][l]{\fboxsep0pt\colorbox{Mywhite} {\strut\hspace*{\linewidth}}}
5 & 69 & 275/14 & 1869 & 91 & 195/8 & 3390 \\
\makebox[0pt][l]{\fboxsep0pt\colorbox{Mygrey} {\strut\hspace*{\linewidth}}}
6 & 119 & 22 & 6013 & 168 & 468/17 & 12420 \\
\makebox[0pt][l]{\fboxsep0pt\colorbox{Mywhite} {\strut\hspace*{\linewidth}}}
7 & 189 & 385/16 & 15897 & 280 & 91/3 & 35940 \\
\makebox[0pt][l]{\fboxsep0pt\colorbox{Mygrey} {\strut\hspace*{\linewidth}}}
8 & 294 & 440/17 & 39711 & 462 & 624/19 & 100515 \\
\makebox[0pt][l]{\fboxsep0pt\colorbox{Mywhite} {\strut\hspace*{\linewidth}}}
9 & 434 & 55/2 & 88375 & 714 & 351/10 & 244155 \\
\makebox[0pt][l]{\fboxsep0pt\colorbox{Mygrey} {\strut\hspace*{\linewidth}}}
10 & 630 & 550/19 & 189063 & 1092 & 260/7 & 577866 \\
\makebox[0pt][l]{\fboxsep0pt\colorbox{Mywhite} {\strut\hspace*{\linewidth}}}
11 & 882 & 121/4 & 374535 & 1596 & 39 & 1244154 \\
\makebox[0pt][l]{\fboxsep0pt\colorbox{Mygrey} {\strut\hspace*{\linewidth}}}
12 & 1218 & 220/7 & 719985 & 2310 & 936/23 & 2621355 \\
\hline
\hline
\end{tabular*} 
\end{threeparttable}
\end{center}
\end{table}

\begin{table}[h]
\begin{center}
\begin{threeparttable}
\caption{C series}
\begin{tabular*}{\textwidth}{@{\extracolsep{\fill}}cccccccccc}
\hline
\hline
\multicolumn{1}{c}{} & \multicolumn{3}{c}{$C_2$} & \multicolumn{3}{c}{$C_3$} & \multicolumn{3}{c}{$C_4$} \\
\cline{2-4} \cline{5-7} \cline{8-10}
\multicolumn{1}{c}{k} & \multicolumn{1}{c}{n} & \multicolumn{1}{c}{c} & \multicolumn{1}{c}{l} & \multicolumn{1}{c}{n} & \multicolumn{1}{c}{c} & \multicolumn{1}{c}{l} & \multicolumn{1}{c}{n} & \multicolumn{1}{c}{c} & \multicolumn{1}{c}{l} \\ 
\hline
\makebox[0pt][l]{\fboxsep0pt\colorbox{Mywhite} {\strut\hspace*{\linewidth}}}
1 & 3 & 5/2 & 0 & 4 & 21/5 & 0 & 5 & 6 & 0  \\
\makebox[0pt][l]{\fboxsep0pt\colorbox{Mygrey} {\strut\hspace*{\linewidth}}}
2 & 6 & 4 & 3 & 10 & 7 & 15 & 15 & 72/7 & 45  \\
\makebox[0pt][l]{\fboxsep0pt\colorbox{Mywhite} {\strut\hspace*{\linewidth}}}
3 & 10 & 5 & 15 & 20 & 9 & 100 & 35 & 27/2 & 385  \\
\makebox[0pt][l]{\fboxsep0pt\colorbox{Mygrey} {\strut\hspace*{\linewidth}}}
4 & 15 & 40/7 & 45 & 35 & 21/2 & 385 & 70 & 16 & 1855  \\
\makebox[0pt][l]{\fboxsep0pt\colorbox{Mywhite} {\strut\hspace*{\linewidth}}}
5 & 21 & 25/4 & 105 & 56 & 35/3 & 1120 & 126 & 18 & 6615  \\
\makebox[0pt][l]{\fboxsep0pt\colorbox{Mygrey} {\strut\hspace*{\linewidth}}}
6 & 28 & 20/3 & 210 & 84 & 63/5 & 2730 & 210 & 216/11 & 19425  \\
\makebox[0pt][l]{\fboxsep0pt\colorbox{Mywhite} {\strut\hspace*{\linewidth}}}
7 & 36 & 7 & 378 & 120 & 147/11 & 5880 & 330 & 21 & 49665  \\
\makebox[0pt][l]{\fboxsep0pt\colorbox{Mygrey} {\strut\hspace*{\linewidth}}}
8 & 45 & 80/11 & 630 & 165 & 14 & 11550 & 495 & 288/13 & 114345  \\
\makebox[0pt][l]{\fboxsep0pt\colorbox{Mywhite} {\strut\hspace*{\linewidth}}}
9 & 55 & 15/2 & 990 & 220 & 189/13 & 21120 & 715 & 162/7 & 242385  \\
\makebox[0pt][l]{\fboxsep0pt\colorbox{Mygrey} {\strut\hspace*{\linewidth}}}
10 & 66 & 100/13 & 1485 & 286 & 15 & 36465 & 1001 & 24 & 480480  \\
\makebox[0pt][l]{\fboxsep0pt\colorbox{Mywhite} {\strut\hspace*{\linewidth}}}
11 & 78 & 55/7 & 2145 & 364 & 77/5 & 60060 & 1365 & 99/4 & 900900  \\
\makebox[0pt][l]{\fboxsep0pt\colorbox{Mygrey} {\strut\hspace*{\linewidth}}}
12 & 91 & 8 & 3003 & 455 & 63/4 & 95095 & 1820 & 432/17 & 1611610  \\
\hline
\hline
\end{tabular*} 
\end{threeparttable}
\end{center}
\end{table}

\begin{table}[h]
\begin{center}
\begin{threeparttable}
\begin{tabular*}{\textwidth}{@{\extracolsep{\fill}}ccccccc}
\hline
\hline
\multicolumn{1}{c}{} & \multicolumn{3}{c}{$C_5$} & \multicolumn{3}{c}{$C_6$} \\
\cline{2-4} \cline{5-7}
\multicolumn{1}{c}{k} & \multicolumn{1}{c}{n} & \multicolumn{1}{c}{c} & \multicolumn{1}{c}{l} & \multicolumn{1}{c}{n} & \multicolumn{1}{c}{c} & \multicolumn{1}{c}{l} \\ 
\hline
\makebox[0pt][l]{\fboxsep0pt\colorbox{Mywhite} {\strut\hspace*{\linewidth}}}
1 & 6 & 55/7 & 0 & 7 & 39/4 & 0  \\
\makebox[0pt][l]{\fboxsep0pt\colorbox{Mygrey} {\strut\hspace*{\linewidth}}}
2 & 21 & 55/4 & 105 & 28 & 52/3 & 210  \\
\makebox[0pt][l]{\fboxsep0pt\colorbox{Mywhite} {\strut\hspace*{\linewidth}}}
3 & 56 & 55/3 & 1120 & 84 & 117/5 & 2730  \\
\makebox[0pt][l]{\fboxsep0pt\colorbox{Mygrey} {\strut\hspace*{\linewidth}}}
4 & 126 & 22 & 6615 & 210 & 312/11 & 19425  \\
\makebox[0pt][l]{\fboxsep0pt\colorbox{Mywhite} {\strut\hspace*{\linewidth}}}
5 & 252 & 25 & 28476 & 462 & 65/2 & 99561  \\
\makebox[0pt][l]{\fboxsep0pt\colorbox{Mygrey} {\strut\hspace*{\linewidth}}}
6 & 462 & 55/2 & 99561 & 924 & 36 & 409794  \\
\makebox[0pt][l]{\fboxsep0pt\colorbox{Mywhite} {\strut\hspace*{\linewidth}}}
7 & 792 & 385/13 & 299376 & 1716 & 39 & 1435434  \\
\makebox[0pt][l]{\fboxsep0pt\colorbox{Mygrey} {\strut\hspace*{\linewidth}}}
8 & 1287 & 220/7 & 801801 & 3003 & 208/5 & 4435431  \\
\makebox[0pt][l]{\fboxsep0pt\colorbox{Mywhite} {\strut\hspace*{\linewidth}}}
9 & 2002 & 33 & 1957956 & 5005 & 351/8 & 12387375  \\
\makebox[0pt][l]{\fboxsep0pt\colorbox{Mygrey} {\strut\hspace*{\linewidth}}}
10 & 3003 & 275/8 & 4432428 & 8008 & 780/17 & 31819788  \\
\makebox[0pt][l]{\fboxsep0pt\colorbox{Mywhite} {\strut\hspace*{\linewidth}}}
11 & 4368 & 605/17 & 9417408 & 12376 & 143/3 & 76168092  \\
\makebox[0pt][l]{\fboxsep0pt\colorbox{Mygrey} {\strut\hspace*{\linewidth}}}
12 & 6188 & 110/3 & 18956938 & 18564 & 936/19 & 171633462  \\
\hline
\hline
\end{tabular*} 
\end{threeparttable}
\end{center}
\end{table}

\begin{table}[h]
\begin{center}
\begin{threeparttable}
\caption{D series}
\begin{tabular*}{\textwidth}{@{\extracolsep{\fill}}ccccccccccccc}
\hline
\hline
\multicolumn{1}{c}{} & \multicolumn{3}{c}{$D_4$} & \multicolumn{3}{c}{$D_5$} & \multicolumn{3}{c}{$D_6$} \\
\cline{2-4} \cline{5-7} \cline{8-10} 
\multicolumn{1}{c}{k} & \multicolumn{1}{c}{n} & \multicolumn{1}{c}{c} & \multicolumn{1}{c}{l} & \multicolumn{1}{c}{n} & \multicolumn{1}{c}{c} & \multicolumn{1}{c}{l} & \multicolumn{1}{c}{n} & \multicolumn{1}{c}{c} & \multicolumn{1}{c}{l} \\ 
\hline
\makebox[0pt][l]{\fboxsep0pt\colorbox{Mywhite} {\strut\hspace*{\linewidth}}}
1 & 2 & 4 & 0 & 3 & 5 & 0 & 3 & 6 & 0 \\
\makebox[0pt][l]{\fboxsep0pt\colorbox{Mygrey} {\strut\hspace*{\linewidth}}}
2 & 5 & 7 & 0 & 9 & 9 & 12 & 10 & 11 & 15 \\
\makebox[0pt][l]{\fboxsep0pt\colorbox{Mywhite} {\strut\hspace*{\linewidth}}}
3 & 9 & 28/3 & 8 & 19 & 135/11 & 96 & 22 & 198/13 & 135 \\
\makebox[0pt][l]{\fboxsep0pt\colorbox{Mygrey} {\strut\hspace*{\linewidth}}}
4 & 16 & 56/5 & 49 & 39 & 15 & 525 & 49 & 132/7 & 870 \\
\makebox[0pt][l]{\fboxsep0pt\colorbox{Mywhite} {\strut\hspace*{\linewidth}}}
5 & 25 & 140/11 & 160 & 69 & 225/13 & 1869 & 91 & 22 & 3390 \\
\makebox[0pt][l]{\fboxsep0pt\colorbox{Mygrey} {\strut\hspace*{\linewidth}}}
6 & 39 & 14 & 469 & 119 & 135/7 & 6013 & 168 & 99/4 & 12420 \\
\makebox[0pt][l]{\fboxsep0pt\colorbox{Mywhite} {\strut\hspace*{\linewidth}}}
7 & 56 & 196/13 & 1083 & 189 & 21 & 15897 & 280 & 462/17 & 35940 \\
\makebox[0pt][l]{\fboxsep0pt\colorbox{Mygrey} {\strut\hspace*{\linewidth}}}
8 & 80 & 16 & 2400 & 294 & 45/2 & 39711 & 462 & 88/3 & 100515 \\
\makebox[0pt][l]{\fboxsep0pt\colorbox{Mywhite} {\strut\hspace*{\linewidth}}}
9 & 109 & 84/5 & 4715 & 434 & 405/17 & 88375 & 714 & 594/19 & 244155 \\
\makebox[0pt][l]{\fboxsep0pt\colorbox{Mygrey} {\strut\hspace*{\linewidth}}}
10 & 147 & 35/2 & 8958 & 630 & 25 & 189063 & 1092 & 33 & 577866 \\
\makebox[0pt][l]{\fboxsep0pt\colorbox{Mywhite} {\strut\hspace*{\linewidth}}}
11 & 192 & 308/17 & 15780 & 882 & 495/19 & 374535 & 1596 & 242/7 & 1244154 \\
\makebox[0pt][l]{\fboxsep0pt\colorbox{Mygrey} {\strut\hspace*{\linewidth}}}
12 & 249 & 56/3 & 27231 & 1218 & 27 & 719985 & 2310 & 36 & 2621355 \\
\hline
\hline
\end{tabular*} 
\end{threeparttable}
\end{center}
\end{table}

\clearpage

\begin{table}[]
\begin{center}
\begin{threeparttable}
\caption{Exceptional series}
\begin{tabular*}{\textwidth}{@{\extracolsep{\fill}}cccccccccc}
\hline
\hline
\multicolumn{1}{c}{} & \multicolumn{3}{c}{$E_6$} & \multicolumn{3}{c}{$E_7$} & \multicolumn{3}{c}{$E_8$} \\
\cline{2-4} \cline{5-7} \cline{8-10}
\multicolumn{1}{c}{k} & \multicolumn{1}{c}{n} & \multicolumn{1}{c}{c} & \multicolumn{1}{c}{l} & \multicolumn{1}{c}{n} & \multicolumn{1}{c}{c} & \multicolumn{1}{c}{l} & \multicolumn{1}{c}{n} & \multicolumn{1}{c}{c} & \multicolumn{1}{c}{l} \\ 
\hline
\makebox[0pt][l]{\fboxsep0pt\colorbox{Mywhite} {\strut\hspace*{\linewidth}}}
1 & 2 & 6 & 0 & 2 & 7 & 0 & 1 & 8 & 2  \\
\makebox[0pt][l]{\fboxsep0pt\colorbox{Mygrey} {\strut\hspace*{\linewidth}}}
2 & 6 & 78/7 & 0 & 6 & 133/10 & 0 & 3 & 31/2 & 0  \\
\makebox[0pt][l]{\fboxsep0pt\colorbox{Mywhite} {\strut\hspace*{\linewidth}}}
3 & 12 & 78/5 & 20 & 12 & 19 & 20 & 5 & 248/11 & 0  \\
\makebox[0pt][l]{\fboxsep0pt\colorbox{Mygrey} {\strut\hspace*{\linewidth}}}
4 & 25 & 39/2 & 160 & 25 & 266/11 & 160 & 10 & 496/17 & 5  \\
\makebox[0pt][l]{\fboxsep0pt\colorbox{Mywhite} {\strut\hspace*{\linewidth}}}
5 & 44 & 390/17 & 636 & 44 & 665/23 & 636 & 15 & 248/7 & 33  \\
\makebox[0pt][l]{\fboxsep0pt\colorbox{Mygrey} {\strut\hspace*{\linewidth}}}
6 & 78 & 26 & 2320 & 79 & 133/4 & 2384 & 27 & 124/3 & 161  \\
\makebox[0pt][l]{\fboxsep0pt\colorbox{Mywhite} {\strut\hspace*{\linewidth}}}
7 & 126 & 546/19 & 6580 & 128 & 931/25 & 6804 & 39 & 1736/37 & 429  \\
\makebox[0pt][l]{\fboxsep0pt\colorbox{Mygrey} {\strut\hspace*{\linewidth}}}
8 & 202 & 156/5 & 17887 & 208 & 532/13 & 19015 & 63 & 992/19 & 1320 \\
\makebox[0pt][l]{\fboxsep0pt\colorbox{Mywhite} {\strut\hspace*{\linewidth}}}
9 & 306 & 234/7 & 42532 & 318 & 133/3 & 46056 & 90 & 744/13 & 2985 \\
\makebox[0pt][l]{\fboxsep0pt\colorbox{Mygrey} {\strut\hspace*{\linewidth}}}
10 & 458 & 390/11 & 97712 & 483 & 95/2 & 108972 & 135 & 62 & 7263 \\
\makebox[0pt][l]{\fboxsep0pt\colorbox{Mywhite} {\strut\hspace*{\linewidth}}}
11 & 660 & 858/23 & 206428 & 704 & 1463/29 & 235488 & 90 & 744/13 & 2985 \\
\makebox[0pt][l]{\fboxsep0pt\colorbox{Mygrey} {\strut\hspace*{\linewidth}}}
12 & 940 & 39 & 424063 & 1019 & 266/5 & 499601 & 135 & 62 & 7263 \\
\hline
\hline
\end{tabular*} 
\end{threeparttable}
\end{center}
\end{table}

\begin{table}[]
\begin{center}
\begin{threeparttable}
\begin{tabular*}{\textwidth}{@{\extracolsep{\fill}}ccccccc}
\hline
\hline
\multicolumn{1}{c}{} & \multicolumn{3}{c}{$F_4$} & \multicolumn{3}{c}{$G_2$} \\
\cline{2-4} \cline{5-7}
\multicolumn{1}{c}{k} & \multicolumn{1}{c}{n} & \multicolumn{1}{c}{c} & \multicolumn{1}{c}{l} & \multicolumn{1}{c}{n} & \multicolumn{1}{c}{c} & \multicolumn{1}{c}{l} \\ 
\hline
\makebox[0pt][l]{\fboxsep0pt\colorbox{Mywhite} {\strut\hspace*{\linewidth}}}
1 & 2 & 26/5 & 0 & 2 & 14/5 & 0  \\
\makebox[0pt][l]{\fboxsep0pt\colorbox{Mygrey} {\strut\hspace*{\linewidth}}}
2 & 5 & 104/11 & 0 & 4 & 14/3 & 0  \\
\makebox[0pt][l]{\fboxsep0pt\colorbox{Mywhite} {\strut\hspace*{\linewidth}}}
3 & 9 & 13 & 8 & 6 & 6 & 2  \\
\makebox[0pt][l]{\fboxsep0pt\colorbox{Mygrey} {\strut\hspace*{\linewidth}}}
4 & 16 & 16 & 49 & 9 & 7 & 8  \\
\makebox[0pt][l]{\fboxsep0pt\colorbox{Mywhite} {\strut\hspace*{\linewidth}}}
5 & 25 & 130/7 & 160 & 12 & 70/9 & 20  \\
\makebox[0pt][l]{\fboxsep0pt\colorbox{Mygrey} {\strut\hspace*{\linewidth}}}
6 & 39 & 104/5 & 469 & 16 & 42/5 & 44  \\
\makebox[0pt][l]{\fboxsep0pt\colorbox{Mywhite} {\strut\hspace*{\linewidth}}}
7 & 56 & 91/4 & 1083 & 20 & 98/11 & 80  \\
\makebox[0pt][l]{\fboxsep0pt\colorbox{Mygrey} {\strut\hspace*{\linewidth}}}
8 & 80 & 416/17 & 2400 & 25 & 28/3 & 140  \\
\makebox[0pt][l]{\fboxsep0pt\colorbox{Mywhite} {\strut\hspace*{\linewidth}}}
9 & 109 & 26 & 4715 & 30 & 126/13 & 220  \\
\makebox[0pt][l]{\fboxsep0pt\colorbox{Mygrey} {\strut\hspace*{\linewidth}}}
10 & 147 & 520/19 & 8958 & 36 & 10 & 340  \\
\makebox[0pt][l]{\fboxsep0pt\colorbox{Mywhite} {\strut\hspace*{\linewidth}}}
11 & 192 & 143/5 & 15780 & 42 & 154/15 & 490  \\
\makebox[0pt][l]{\fboxsep0pt\colorbox{Mygrey} {\strut\hspace*{\linewidth}}}
12 & 249 & 208/7 & 27231 & 49 & 21/2 & 700  \\
\hline
\hline
\end{tabular*} 
\begin{tablenotes}
\small
\item k: level of the theory
\item n: number of linearly independent characters of the theory
\item c: central charge of the theory
\item l: Wronksian index of the theory
\end{tablenotes}
\end{threeparttable}
\end{center}
\end{table}

\clearpage

\section{Simple Lie Algebras}\label{app2}
Here we present some properties of all the simple Lie Algebras, which is sourced from the appendix of chapter $13$ of \cite{DiFrancesco:1997nk}: $A_r$, $B_r$, $C_r$, $D_r$, $E_6$, $E_7$, $E_8$, $F_4$ and $G_2$. We use the Cartan notation and for all the classical Lie Algebras, the compact real form is also presented in the parentheses. For every algebra, we give its Dynkin diagram, its dimension ($D$), its Dual Coxeter number ($g$) and its associated Quadratic Form Matrix ($F$). Short roots are represented by black nodes in a Dynkin diagram. The numbers which appear beside the nodes of a Dynkin diagram correspond to the numbering of the corresponding simple root, its mark and its co-mark respectively in this order. For simply laced algebras, we have that the marks and comarks are identical and hence the third entry is omitted for these algebras. The numbering of the simple roots also indicates the numbering of the fundamental weights in the sense that this is the numbering used when a weight is specified in terms of a sequence of Dynkin labels as, $\lambda$ = ($\lambda_1,\cdots\cdots,\lambda_r$). 

\subsection*{$\mathbf{A_{r\geq 1}}(su(r+1))$}
\begin{itemize}
\item Dynkin Diagram:\\
\\
\begin{dynkinDiagram}[text style={scale=1.2,blue},
edge length=25mm,
labels={(1;1),(2;1),(r;1)},
]A{o2.o}
\end{dynkinDiagram}
\\
\item Dimension and dual Coxeter number:\\
\begin{align}\label{2aDG}
&D = r^2+2r \nonumber\\
&g = r+1 
\end{align}
\item Quadratic Form Matrix:
\begin{align}\label{2aQ}
F \ = \ \frac{1}{r+1}\begin{pmatrix}
r & r-1 & r-2 & \cdots & \cdots & \cdots & 2 & 1\\
r-1 & 2(r-1) & 2(r-2) & \cdots & \cdots & \cdots & 4 & 2\\
r-2 & 2(r-2) & 3(r-2) & \cdots & \cdots & \cdots & 6 & 3\\
\vdots & \vdots & \vdots & \vdots & \vdots & \vdots & \vdots & \vdots\\
r-m+1 & 2(r-m+1) & 3(r-m+1) & \cdots & m(r-m+1) & \cdots & 2m & m\\ 
\vdots & \vdots & \vdots & \vdots & \vdots & \vdots & \cdots & \vdots\\
2 & 4 & 6 & \cdots & \cdots & \cdots & 2(r-1) & r-1\\
1 & 2 & 3 & \cdots & \cdots & \cdots & r-1 & r
\end{pmatrix}
\end{align}
\end{itemize}

\subsection*{$\mathbf{B_{r\geq 3}}(so(2r+1))$}
\begin{itemize}
\item Dynkin Diagram:\\
\\
\begin{dynkinDiagram}[text style={scale=1.2,blue},
edge length=25mm,
labels={(1;1;1),(2;2;2),(r-1;2;2),(r;2;1)},
]B{o2.o*}
\end{dynkinDiagram}
\\
\item Dimension and dual Coxeter number:\\
\begin{align}\label{2bDG}
&D = 2r^2+r \nonumber\\
&g = 2r-1
\end{align}
\item Quadratic Form Matrix: 
\begin{align}\label{2bQ}
F \ = \ \frac{1}{2}\begin{pmatrix}
2 & 2 & 2 & \cdots & \cdots & \cdots & 2 & 1\\
2 & 4 & 4 & \cdots & \cdots & \cdots & 4 & 2\\
2 & 4 & 6 & \cdots & \cdots & \cdots & 6 & 3\\
\vdots & \vdots & \vdots & \vdots & \vdots & \vdots & \vdots & \vdots\\
2 & 4 & 6 & \cdots & 2m & \cdots & 2m & m\\
\vdots & \vdots & \vdots & \vdots & \vdots & \vdots & \vdots & \vdots\\
2 & 4 & 6 & \cdots & \cdots & \cdots  & 2(r-1) & r-1\\
1 & 2 & 3 & \cdots & \cdots & \cdots  & r-1 & r/2
\end{pmatrix}
\end{align}
\end{itemize}

\subsection*{$\mathbf{C_{r\geq 2}}(sp(2r))$}
\begin{itemize}
\item Dynkin Diagram:\\
\\
\begin{dynkinDiagram}[text style={scale=1.2,blue},
edge length=25mm,
labels={(1;2;1),(2;2;1),(r-1;2;1),(r;1;1)},
]C{*2.*o}
\end{dynkinDiagram}
\\
\item Dimension and dual Coxeter number:\\
\begin{align}\label{2cDG}
&D = 2r^2+r \nonumber\\
&g = r+1
\end{align}
\item Quadratic Form Matrix:
\begin{align}\label{2cQ}
F \ = \ \frac{1}{2}\begin{pmatrix}
1 & 1 & 1 & \cdots & \cdots & \cdots & 1 & 1\\
1 & 2 & 2 & \cdots & \cdots & \cdots & 2 & 2\\
1 & 2 & 3 & \cdots & \cdots & \cdots & 3 & 3\\
\vdots & \vdots & \vdots & \vdots & \vdots & \vdots & \vdots & \vdots\\
1 & 2 & 3 & \cdots & m & \cdots & m & m\\
\vdots & \vdots & \vdots & \vdots & \vdots & \vdots & \vdots & \vdots\\
1 & 2 & 3 & \cdots & \cdots & \cdots & r-1 & r-1\\
1 & 2 & 3 & \cdots & \cdots & \cdots & r-1 & r
\end{pmatrix}
\end{align}
\end{itemize}

\subsubsection*{$\mathbf{D_{r\geq 4}}(so(2r))$}
\begin{itemize}
\item Dynkin Diagram:\\
\\
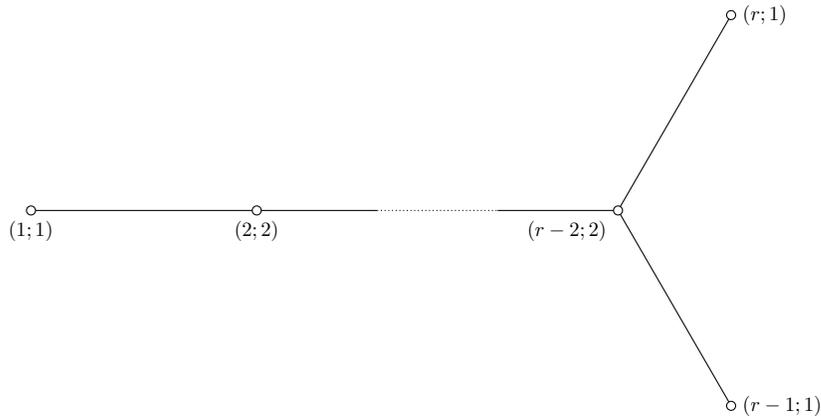
\begin{dynkinDiagram}[text style={scale=1.2,blue},
edge length=25mm,
labels={(1;1),(2;2),(r-2;2),(r;1),(r-1;1)},
]D{o2.o3}
\end{dynkinDiagram}
\\
\item Dimension and dual Coxeter number:\\
\begin{align}\label{2dDG}
&D = 2r^2-r \nonumber\\
&g = 2r-2
\end{align}
\item Quadratic Form Matrix:
\begin{align}\label{2dQ}
F \ = \ \frac{1}{2}\begin{pmatrix}
2 & 2 & 2 & \cdots & \cdots & \cdots & 2 & 1 & 1\\
2 & 4 & 4 & \cdots & \cdots & \cdots & 4 & 2 & 2\\
2 & 4 & 6 & \cdots & \cdots & \cdots & 6 & 3 & 3\\
\vdots & \vdots & \vdots & \vdots & \vdots & \vdots & \vdots & \vdots & \vdots\\
2 & 4 & 6 & \cdots & 2m & \cdots & 2m & m & m\\
\vdots & \vdots & \vdots & \vdots & \vdots & \vdots & \vdots & \vdots & \vdots\\
2 & 4 & 6 & \cdots & \cdots & \cdots & 2(r-2) & r-2 & r-2\\
1 & 2 & 3 & \cdots & \cdots & \cdots & r-2 & r/2 & (r-2)/2\\
1 & 2 & 3 & \cdots & \cdots & \cdots & r-2 & (r-2)/2 & r/2
\end{pmatrix}
\end{align}
\end{itemize}

\subsubsection*{$\mathbf{E_6}$}
\begin{itemize}
\item Dynkin Diagram:\\
\\
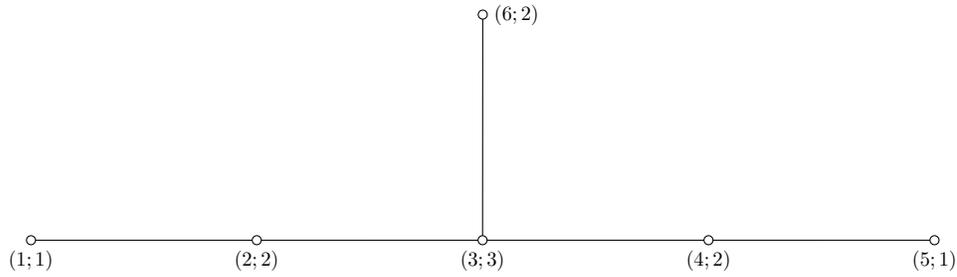
\begin{dynkinDiagram}[text style={scale=1.2,blue},
edge length=25mm,
labels={(1;1),(6;2),(2;2),(3;3),(4;2),(5;1)},mark=o
]E6
\end{dynkinDiagram}
\\
\item Dimension and dual Coxeter number:\\
\begin{align}\label{2e6DG}
&D = 78 \nonumber\\
&g = 12
\end{align}
\item Quadratic Form Matrix:
\begin{align}\label{2e6Q}
F \ = \ \frac{1}{3}\begin{pmatrix}
4 & 5 & 6 & 4 & 2 & 3\\
5 & 10 & 12 & 8 & 4 & 6\\
6 & 12 & 18 & 12 & 6 & 9\\
4 & 8 & 12 & 10 & 5 & 6\\
2 & 4 & 6 & 5 & 4 & 3\\
3 & 6 & 9 & 6 & 3 & 6
\end{pmatrix}
\end{align}
\end{itemize}

\subsubsection*{$\mathbf{E_7}$}
\begin{itemize}
\item Dynkin Diagram:\\
\\
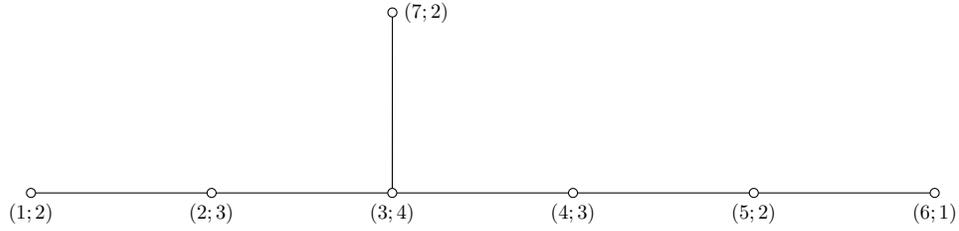
\begin{dynkinDiagram}[text style={scale=1.2,blue},
edge length=20mm,
labels={(1;2),(7;2),(2;3),(3;4),(4;3),(5;2),(6;1)},mark=o
]E7
\end{dynkinDiagram}
\\
\item Dimension and dual Coxeter number:\\
\begin{align}\label{2e7DG}
&D = 133 \nonumber\\
&g = 18
\end{align}
\item Quadratic Form Matrix: 
\begin{align}\label{2e7Q}
F \ = \ \frac{1}{2}\begin{pmatrix}
4 & 6 & 8 & 6 & 4 & 2 & 4\\
6 & 12 & 16 & 12 & 8 & 4 & 8\\
8 & 16 & 24 & 18 & 12 & 6 & 12\\
6 & 12 & 18 & 15 & 10 & 5 & 9\\
4 & 8 & 12 & 10 & 8 & 4 & 6\\
2 & 4 & 6 & 5 & 4 & 3 & 3\\
4 & 8 & 12 & 9 & 6 & 3 & 7
\end{pmatrix}
\end{align}
\end{itemize}

\subsubsection*{$\mathbf{E_8}$}
\begin{itemize}
\item Dynkin Diagram:\\
\\
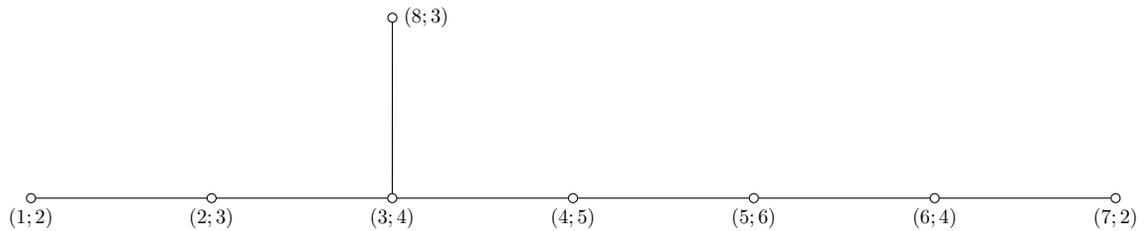
\begin{dynkinDiagram}[text style={scale=1.2,blue},
edge length=20mm,
labels={(1;2),(8;3),(2;3),(3;4),(4;5),(5;6),(6;4),(7;2)},mark=o
]E8
\end{dynkinDiagram}
\\
\item Dimension and Dual Coxeter number:\\
\begin{align}\label{2e8DG}
&D = 248 \nonumber\\
&g = 30
\end{align}
\item Quadratic Form Matrix:
\begin{align}\label{2e8Q}
F \ = \ \begin{pmatrix}
2 & 3 & 4 & 5 & 6 & 4 & 2 & 3\\
3 & 6 & 8 & 10 & 12 & 8 & 4 & 6\\
4 & 8 & 12 & 15 & 18 & 12 & 6 & 9\\
5 & 10 & 15 & 20 & 24 & 16 & 8 & 12\\
6 & 12 & 18 & 24 & 30 & 20 & 10 & 15\\
4 & 8 & 12 & 16 & 20 & 14 & 7 & 10\\
2 & 4 & 6 & 8 & 10 & 7 & 4 & 5\\
3 & 6 & 9 & 12 & 15 & 10 & 5 & 8
\end{pmatrix}
\end{align}
\end{itemize}

\subsubsection*{$\mathbf{F_4}$}
\begin{itemize}
\item Dynkin Diagram:\\
\\
\begin{dynkinDiagram}[text style={scale=1.2,blue},
edge length=30mm,
labels={(1;2;2),(2;3;3),(3;4;2),(4;2;1)},
]F4
\dynkinRootMark{o}1
\dynkinRootMark{o}2
\dynkinRootMark{*}3
\dynkinRootMark{*}4
\end{dynkinDiagram}
\\
\item Dimension and Dual Coxeter number:\\
\begin{align}\label{2f4DG}
&D = 52 \nonumber\\
&g = 9
\end{align}
\item Quadratic Form Matrix:
\begin{align}\label{2f4Q}
F \ = \ \begin{pmatrix}
2 & 3 & 2 & 1\\
3 & 6 & 4 & 2\\
2 & 4 & 3 & 3/2\\
1 & 2 & 3/2 & 1\\
\end{pmatrix}
\end{align}
\end{itemize}

\subsubsection*{$\mathbf{G_2}$}
\begin{itemize}
\item Dynkin Diagram:\\
\\
\begin{dynkinDiagram}[text style={scale=1.2,blue},
edge length=20mm,
labels={(1;2;2),(2;3;1)},
]G2
\dynkinRootMark{o}1
\dynkinRootMark{*}2
\end{dynkinDiagram}
\\
\item Dimension and Dual Coxeter number:\\
\begin{align}\label{2g2DG}
&D = 14 \nonumber\\
&g = 4
\end{align}
\item Quadratic Form Matrix:
\begin{align}\label{2g2Q}
F \ = \ \begin{pmatrix}
2/3 & 1\\
1 & 2
\end{pmatrix}
\end{align}
\end{itemize}

\end{document}